\documentclass[reprint,10pt,twocolumn,superscriptaddress,nofootinbib]{revtex4-1}

\usepackage[utf8]{inputenc}
\usepackage{graphicx}
\usepackage{bm}
\usepackage[colorlinks=true]{hyperref} %hyperlinks

\hypersetup{
    colorlinks=true,
    linkcolor=blue,
    filecolor=magenta,      
    urlcolor=cyan,
}
\usepackage{amsmath,amssymb,amsbsy}
\usepackage{bbm,mathrsfs,yfonts,eufrak}
\usepackage[usenames,dvipsnames]{color}

\newcommand{\norm}[1]{\ensuremath{|| #1 ||}}

\newcommand{\avgR}[1]{\ensuremath{\langle #1 \rangle}}
\newcommand{\flucRrel}[1]{\ensuremath{\delta{ #1} }}

\renewcommand{\ne}{\ensuremath{n }}
\newcommand{\neref}{\ensuremath{n_0}}
\newcommand{\Ni}{\ensuremath{N}}

\newcommand{\OmegaciN}{\ensuremath{\Omega_0}} %\Omega_{i0}
\newcommand{\rhoN}{\ensuremath{\rho_{s0}}} %\rhoN
\newcommand{\teN}{\ensuremath{T_{e0}}} %\rhoN

\renewcommand{\vec}[1]{{\mathbf{#1}}}
\newcommand{\bhat}{\hat{\vec b}}

\begin{document}
\title{The collisional drift wave instability in steep density gradient regimes}
\author{M.\ Held}
\email[E-mail: ]{markus.held@uibk.ac.at}
\affiliation{Institute for Ion Physics and Applied Physics, 
                     Universit\"at Innsbruck, A-6020 Innsbruck, Austria}
\author{M.\ Wiesenberger}
\affiliation{Department of Physics, Technical University of Denmark, DK-2800 Kgs. Lyngby, Denmark}
\author{A.\ Kendl}
\affiliation{Institute for Ion Physics and Applied Physics, 
                     Universit\"at Innsbruck, A-6020 Innsbruck, Austria}
\date{\today}
\begin{abstract} %<600 characters including spaces
The collisional drift wave instability in a straight magnetic field configuration is studied within a full-F gyro-fluid model, 
which relaxes the Oberbeck-Boussinesq (OB) approximation. 
Accordingly, we focus our study on steep background density gradients.
In this regime we report on corrections by factors of order one to the eigenvalue analysis of former OB approximated approaches
as well as on spatially localised eigenfunctions, that contrast strongly with their OB approximated equivalent.
Remarkably, non-modal phenomena arise for large density inhomogeneities and for all collisionalities. 
As a result, we find initial decay and non-modal growth of the free energy and radially localised and sheared growth patterns. 
The latter non-modal effect sustains even in the nonlinear regime in the form of radially localised turbulence or zonal flow amplitudes.
\end{abstract}
\maketitle
\section{Introduction}
In a series of seminal theoretical works in the early 1960s it has been established that low-frequency small scale instabilites are naturally immanent to magnetically confined plasmas~\cite{rudakov61,chen63,chen64,moiseev63,galeev63,krall65}. This is due to the inherent plasma pressure gradients, which nurture these so called drift wave (DW) instabilities.
These DW instabilities drive the turbulent cross-field transport of particles and heat, which 
exceeds predictions from classical and neo-classical theory and remains a serious barrier for sufficient plasma confinement in laboratory plasmas. 

%Physical mechanism and coll, collless DWs in slab, shear and non-modal effects
Unstable DWs are triggered by the non-adiabatic coupling of plasma density fluctuations and the electric potential, which can arise due to various physical mechanisms~\cite{kadomtsev70,mikhailovskii74,tang78,horton99,bellan06}. 
Two of particular importance are (i) collisional friction of electrons and ions along the magnetic field line and (ii) wave-particle resonances. 
The first of these mechanisms is associated to the collisional (resistive, dissipative) DW instability~\cite{chen63,chen64,moiseev63}, on which we focus in this contribution.
The second purely kinetic mechanism results in the so called collisionless (universal) DW instability~\cite{galeev63,krall65}. The latter two DW instabilities may 
be linearly stabilised by magnetic shear~\cite{guzdar78,chen79,pearlstein69,tsang78,ross78, chen78, antonsen78,landreman15,helander15}. 
However, the inherent non-modal character of density gradient driven DWs (cf.~\cite{camargo98,mikhailenko00,friedman15,landreman15}) allows for transient amplification to sustain DW turbulence~\cite{scott90, scott92, drake95,landreman15b}.

%Recent research
After nearly 60 years  density gradient driven DW instabilities remain an active theoretical research field. 
The latest efforts focus on a unified description of the collisional and collisionless DW instability~\cite{angus12,jorge18} or proof instability for collisionless DWs in sheared magnetic fields after decades of misconception~\cite{landreman15}.

Another outstanding regime of interest is that of large inhomogeneities, in particular if the background density varies over more than one order of magnitude. 
This for example prevails in tokamak fusion plasmas, where steep background density gradients can emerge due to the formation of internal or edge transport barriers~\cite{shao16,kobayashi16}.
Under these circumstances the Oberbeck-Boussinesq (OB) approximation~\cite{oberbeck1879,boussinesq03} breaks down, which is throughout applied in former studies of DW instabilites. 
Thus, a rigorous stability analysis for large inhomogeneities must relax the latter assumption.

In spite of that, recent OB approximated analysis of large inhomogeneity effects on trapped electron or ion temperature gradient driven DW instabilities indicate 
their relevance for transport bifurcation in the edge of tokamaks~\cite{xie15,xie17,chen18}. 
Initial investigations of large density inhomogeneity effects on the stability and dynamics of collisional DWs rest partly upon the OB approximation and exploit further approximations in their linear analysis~\cite{mattor94, zhang17}. 

%Outline
In this contribution we investigate the linear dynamics of the collisional DW instability for steep background density gradients.
This is achieved by consistently linearising a non-Oberbeck-Boussinesq (NOB) approximated full-F gyro-fluid model~\cite{madsen13}, 
which accurately accounts for collisional friction between electrons and ions along the magnetic field. 
We numerically solve the generalised eigenvalue problem to show that the growth rate and real frequency deviate by factors of order one 
from the OB approximated case in the steep background density gradient regime. 
In this regime, the NOB approximated model is strongly non-normal. As a consequence the eigenfunctions are spatially localised, as opposed to the OB limit. Moreover, non-modal effects arise,
resulting in initial decay and transient non-modal growth of the free energy
and radially localised and sheared growth of an initially unstable random perturbation.
The latter NOB signature lingers into the nonlinear regime, where radially localised turbulence amplitudes appear.
%gyrofluid model
\section{Gyro-fluid model}
Our  analysis is based on an energetically consistent full-F gyro-fluid model~\cite{madsen13}, which is derived by taking the gyro-fluid moments over the gyro-kinetic 
Vlasov-Maxwell equations~\cite{brizard07}.
In order to ease the following discussion we assume constant temperatures, cold ions and  a constant magnetic field \(B=B_0\) with straight and unsheared unit vector
\(\bhat := \vec{B}/B = \vec{\hat{e}}_z\).
The resulting set of gyro-fluid equations consists of continuity equations for electron density \({\ne}\) and  ion gyro-center density  \(\Ni\) and the 
quasi-neutrality constraint
\begin{subequations} 
\begin{eqnarray}
\label{eq:fullFdtne3d}
 \frac{\partial}{\partial t }\ne  +\vec{\nabla} \cdot \left( \ne \vec{u}_E\right)  &=& \frac{\teN}{ \eta_\parallel  e^2}\nabla_\parallel^2 \left(\ln\left(\ne\right) -\frac{e}{\teN}\phi\right) , \\
\label{eq:fullFdtNi}
\frac{\partial}{\partial t } \Ni +\vec{\nabla} \cdot \left( \Ni \vec{U}_E\right) &=&0 , \\
\label{eq:ffpolcoldlwl}
 \vec{\nabla} \cdot \left(\frac{\Ni}{\OmegaciN} \frac{\vec{\nabla}_\perp \phi}{B_0}\right)&=& \ne - \Ni, 
\end{eqnarray}
\end{subequations}
where \(\phi\) is the electric potential, \(\OmegaciN :=  e B/m_i\) is the ion gyro-frequency and 
\(\vec{\nabla}_\perp := - \bhat \times (\bhat \times \vec{\nabla})\) and \(\nabla_\parallel := \bhat \cdot \vec{\nabla}\) are the perpendicular and parallel gradient, respectively.
The \(\vec{E}\times\vec{B}\) drift velocity is defined by \(\vec{u}_E:= \bhat \times \vec{\nabla} \phi/B_0\).
As opposed to this, the gyro-center \(\vec{E}\times\vec{B}\) drift velocity 
\(\vec{U}_E:= \vec{u}_E + \vec{U}_p\) contains the ponderomotive correction \( \vec{U}_p :=-\bhat \times \vec{\nabla} \vec{u}_E^2/(2 \OmegaciN)\). 

%collisional friction closure
Parallel collisional friction between electrons and ions  is introduced on the right hand side of Eq.~\eqref{eq:fullFdtne3d} 
with  parallel Spitzer resistivity \(\eta_\parallel := 0.51 m_e \nu_e/(\ne e^2) \)~\cite{spitzer56,lingam17}. 
These closures of the Hasegawa-Wakatani (HW) type are obtained from the evolution equation for the parallel electron velocity. 
In the electron collision frequency \(\nu_e :=  \ne  e^4 \ln(\Lambda) /(3 \sqrt{(2 \pi)^{3} \epsilon_0^4 m_e T_e^{3}})\) the Coulomb logarithm \(\ln(\Lambda)\) is treated as a constant~\cite{wesson07}. This is a reasonable 
approximation even if the density profile varies over several orders of magnitude. 
Consequently, the parallel Spitzer resistivity \(\eta_\parallel\) has no explicit dependence on the electron density \(\ne\), since we only retain the electron density \(\ne\) proportionality in the electron collision frequency \(\nu_e \).

%%%%%%Nonlinear Quasi-2D GF model
The two dimensional form of the presented full-F extension of the ordinary HW (OHW) model is obtained by 
rewriting \(\nabla_\parallel^2 \ln\left(\ne\right) \) 
to 
\(\nabla_\parallel^2 \ln\left(\ne/\avgR{\ne}\right) \) and by replacing the parallel derivative with the characteristic parallel wave-number, so that 
\(\nabla_\parallel^2 = - k_\parallel^2\). Here,  the average over the ``poloidal'' y coordinate \(\avgR{  h} := L_y^{-1}\int_0^{L_y} dy \hspace{1mm} h\) is introduced, 
which is the 2D equivalent  of a flux surface average. 
With these manipulations Eq.~\eqref{eq:fullFdtne3d} reduces to~\cite{held18} 
\begin{eqnarray}
\label{eq:fullFdtne}
 \frac{\partial}{\partial t }\ne  +\vec{\nabla} \cdot \left( \ne \vec{u}_E\right) &=\alpha \neref \OmegaciN \left(\frac{e}{\teN}\phi -\ln\left(\ne/\avgR{  \ne }\right) \right),
\end{eqnarray}
where the adiabaticity parameter is 
\begin{eqnarray}
 \alpha:= \frac{ \teN  k_\parallel^2}{\eta_\parallel e^2   \neref \OmegaciN}.
\end{eqnarray}
%validity of the model
The  nonlinear gyro-fluid model of Eqs.~\eqref{eq:fullFdtne},~\eqref{eq:fullFdtNi} and~\eqref{eq:ffpolcoldlwl}  allows to study NOB effects on collisional DWs, since in this regime of large collisionality \(\nu_e \gg \omega\) and small Knudsen-number \(\textit{Kn}:=k_\parallel \lambda_{mfp} \ll 1\) the presented fluid approach is valid.

%%%%%%%Linearised GF Model
\section{Linearised gyro-fluid model}
The linearised gyro-fluid model is obtained by expanding the nonlinear gyro-fluid model of Eqs.~\eqref{eq:fullFdtne},~\eqref{eq:fullFdtNi} and~\eqref{eq:ffpolcoldlwl} 
around a reference background density profile \(n_G(x)\) according to
\begin{eqnarray}
 \ne:= n_G (1+\flucRrel{\ne}), \qquad \Ni:= n_G (1+\flucRrel{\Ni}).
\end{eqnarray}
The chosen exponential reference background density profile \(n_G\)  yields a constant density gradient (e-folding) length  
\begin{eqnarray}
 L_n:= -\left(\frac{\partial}{\partial_x} \ln{(n_G(x)/\neref)}\right)^{-1},
\end{eqnarray}
as in \(\flucRrel{f}\) theory. 
Note that we found similar trends in our results for non-exponential reference background density profiles. 
Assuming that the relative fluctuation amplitudes \(\flucRrel{\ne}\propto\flucRrel{\Ni}\ll1\) are small and the averaged and reference background density profiles coincide \(\avgR{\ne}\approx n_G\) 
 yields the final form of the linearised gyro-fluid model
\begin{subequations} 
\begin{eqnarray}
\label{eq:fullFdtnelin}
 \frac{\partial}{\partial t }\flucRrel{\ne}  +\frac{1}{B_0}\frac{1}{L_n}\frac{\partial }{\partial y}\phi&=&\frac{\alpha \neref \OmegaciN }{n_G}  \left(\frac{e}{\teN}\phi -\flucRrel{\ne}\right),\\
\label{eq:fullFdtNilin}
\frac{\partial}{\partial t } \flucRrel{\Ni} +\frac{1}{B_0}\frac{1}{L_n}\frac{\partial }{\partial y}\phi &=&0, \\
\label{eq:ffpolcoldlwllin}
 -\frac{1}{L_n}\frac{\partial}{\partial x} \phi + \vec{\nabla}_\perp^2 \phi&=& \OmegaciN B_0 \left(\flucRrel{\ne} - \flucRrel{\Ni}\right).
\end{eqnarray}
\end{subequations} 
As opposed to previously exploited linearised models~\cite{mattor94,camargo95, numata07,angus12, zhang17} we do not apply the OB approximation 
in the linearised Eqs.~\eqref{eq:fullFdtnelin},~\eqref{eq:fullFdtNilin} and~\eqref{eq:ffpolcoldlwllin} or in the further course of the calculation.
Thus the derived set of linearised Eqs.~\eqref{eq:fullFdtnelin},~\eqref{eq:fullFdtNilin} and~\eqref{eq:ffpolcoldlwllin} differs from  linearised OB approximated models in two substantial aspects. 
First, we retain the background density on the right hand side of Eq.~\eqref{eq:fullFdtnelin} instead of assuming 
a constant reference density \(n_G \approx n_0\).
Secondly, we preserve the term  \( -\frac{1}{L_n}\frac{\partial}{\partial x} \phi\) on the left hand side of Eq.~\eqref{eq:ffpolcoldlwllin}, which 
originates from the nonlinear contribution of the polarization charge density.
Both of these terms, but especially the NOB approximated resistive term, produce novel linear effects for collisional DWs in steep background density gradients as is shown in section~\ref{sec:lindyn}.
\section{Linear effects}\label{sec:lindyn}
\begin{figure}[ht]
\centering
\includegraphics[width= 0.238\textwidth]{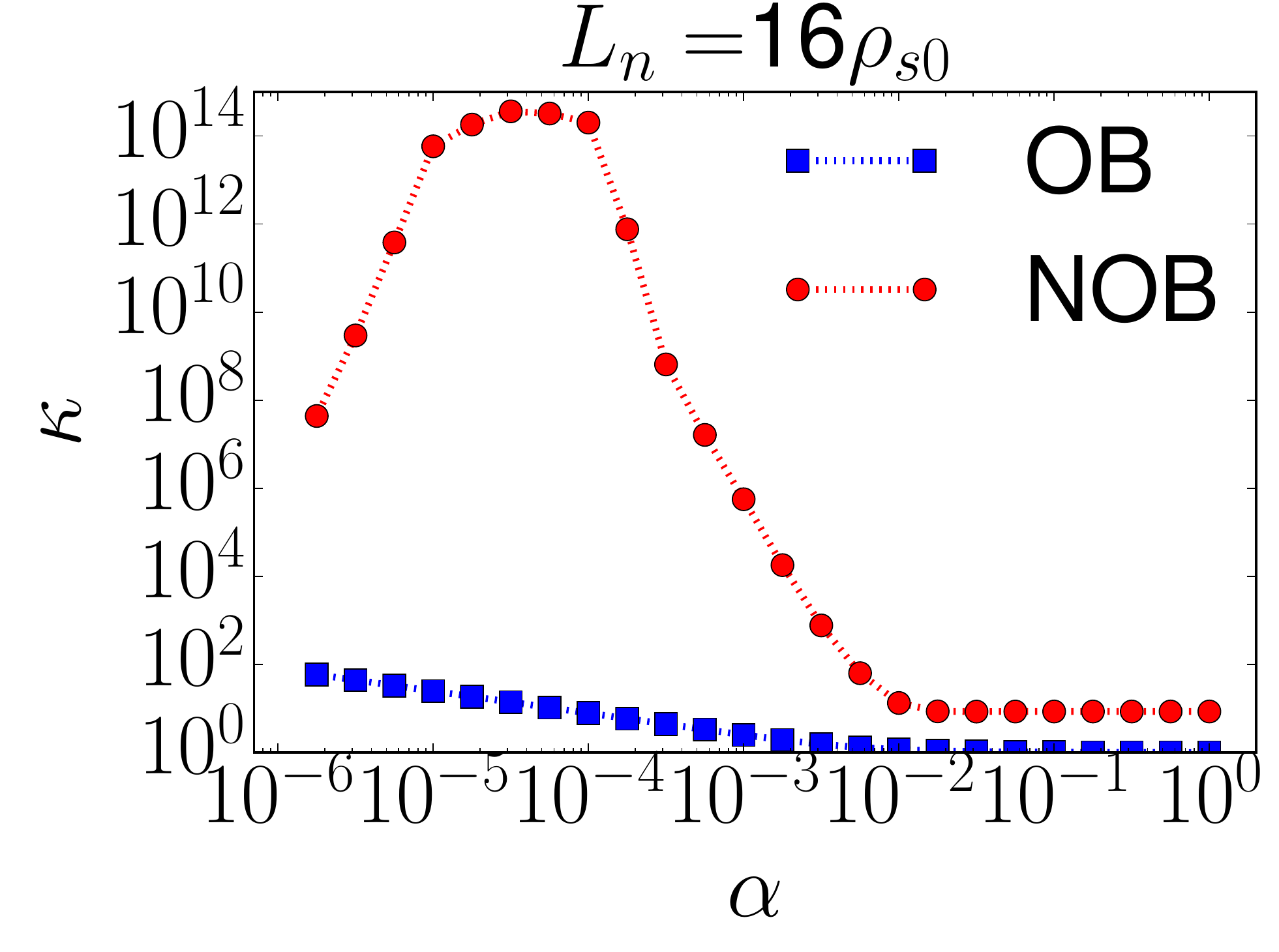}
\includegraphics[width= 0.238\textwidth]{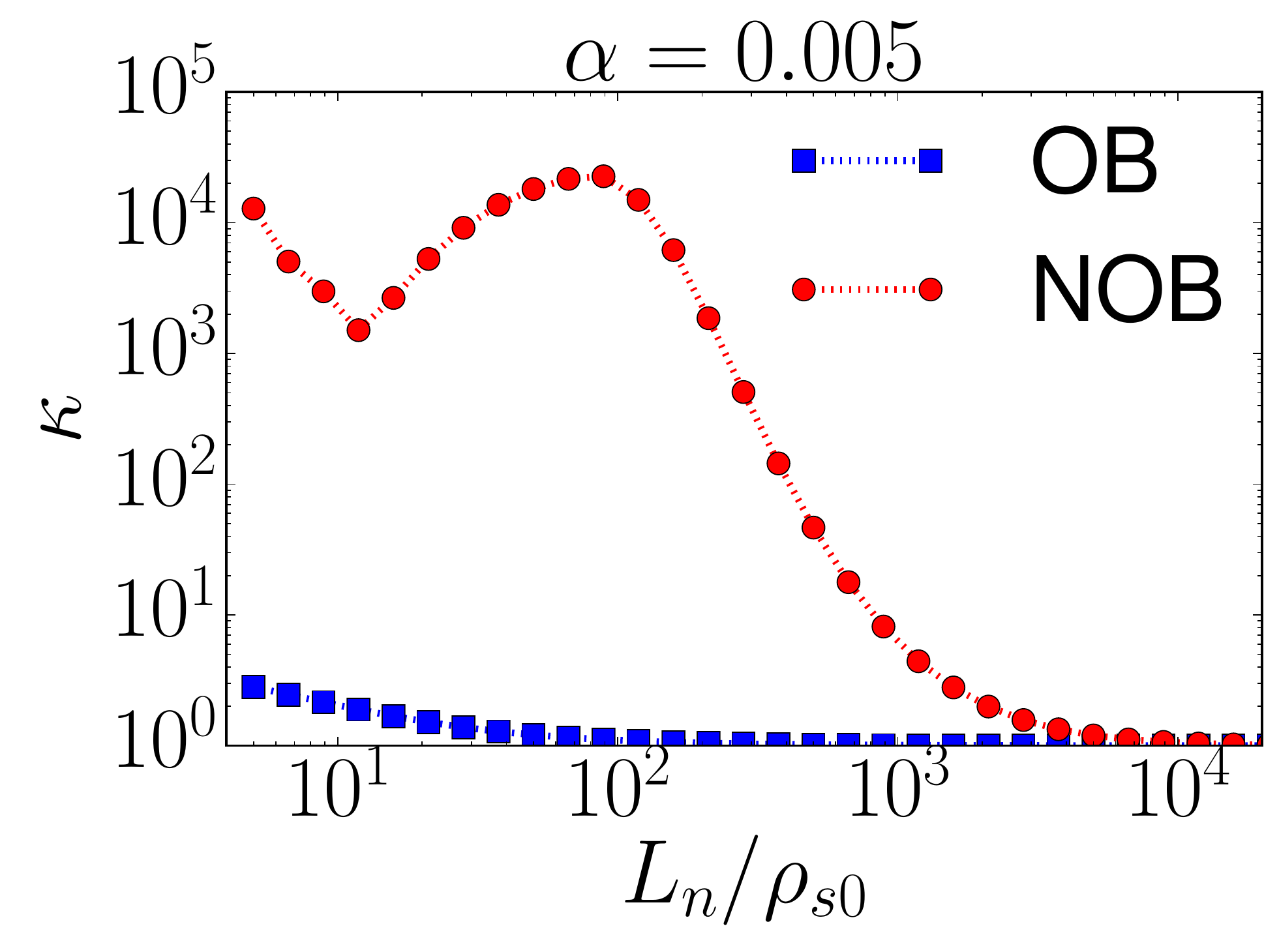}
\caption{The condition number \(\kappa\) is shown for varying adiabaticity \(\alpha\) (left) or background density length \(L_n\) (right). The poloidal wave-number is 
\(k_y=4 \pi /L_y\) and the box size is \(L_x=L_y=64 \rhoN\). Here, we utilise the drift scale \(\rhoN:=\sqrt{\teN m_i }/(e B_0)\) for normalization.
}
\label{fig:nonnormality}
\end{figure}

In the following we want to gain insight into the linear dynamics, in particular in stability and transient time behavior, of the linear model Eqs.~\eqref{eq:fullFdtnelin},~\eqref{eq:fullFdtNilin} and~\eqref{eq:ffpolcoldlwllin}. 
This is accomplished within a discrete approach, which utilises a Fourier transformation of the form \( e^{i  k_y y }\) in the periodic poloidal coordinate \(y\)
and a Galerkin approach with a sine basis in radial direction. This fulfills the chosen Dirichlet boundary conditions in radial direction.
The resulting linear equation reads in matrix form
\begin{eqnarray}\label{eq:lindiscrete}
    \frac{\partial}{\partial t}\vec{v} = \vec{A} \vec{v} ,
\end{eqnarray}
with vector  \(\vec{v}:=(\vec{\flucRrel{\ne}}_{k_y},e \vec{\phi}_{k_y}/\teN)^T\) and 
matrix 
\begin{eqnarray}\label{eq:matrixA}
\vec{A}&:=
\left(\begin{array}{c c}
-\alpha \vec{E}& \alpha\vec{E} -  i \frac{ \rhoN^2 k_y}{L_n }\vec{I} \\
-\alpha \vec{F}^{-1}\vec{E} & \alpha \vec{F}^{-1}\vec{E} \\
\end{array}\right),
\end{eqnarray}
with matrix \(\vec{F}:= \rhoN^2\left( \vec{D}_{xx} - k_y^2 \vec{I}  - \frac{ 1}{ L_n}\vec{D}_x\right)\). 
The coefficients of the symmetric matrix \(\vec{E}\) and \(\vec{D}_{xx}\) and 
of the skew-symmetric matrix \(\vec{D}_x\) are derived to
\begin{eqnarray}
% \qquad
 E_{k_x,k_x'}
 &=&
 \frac{4(e^{L_x/L_n}(-1)^{\frac{(k_x+k_x')L_x}{\pi}}-1) k_x k_x'}{L_x\left[(k_x^2-k_x'^2)^2 L_n + 2 (k_x^2 + k_x'^2)/L_n+1/L_n^3\right]}\nonumber \\
 D_{xx,k_x,k_x'}
 &=&- k_x^2  \delta_{k_x,k_x'}\nonumber \\
 D_{x ,k_x,k_x'}
 &=& 
%  \cases{
%  - \frac{4  k_x k_x'}{L_x(k_x^2 -k_x'^2)}, \qquad \frac{L_x (k_x + k_x')}{\pi}= odd \\
%  0, \qquad  \qquad  \qquad  \qquad else}
  \left\{ \begin{array}{l@{\quad}l} 
- \frac{4  k_x k_x'}{L_x(k_x^2 -k_x'^2)} & \frac{L_x (k_x + k_x')}{\pi}= odd  \\  
0 &   else
\end{array}\right.
\nonumber
\end{eqnarray}
with radial box size \(L_x\). The radial wave-numbers are defined by \(k_x:=\pi m/L_x\) and \(k_x':=\pi m'/L_x\) with mode-numbers \(m \in \mathbb{N}\) and \(m' \in \mathbb{N}\), respectively.\\
%define non-normality
Note, that the matrix \(\vec{A}\) of Eq.~\eqref{eq:matrixA} is in general far from normal (\(\vec{A}\vec{A}^\dagger \neq \vec{A}^\dagger\vec{A}\)) with non-orthogonal eigenvectors 
but approaches a normal matrix (\(\vec{A}\vec{A}^\dagger = \vec{A}^\dagger\vec{A}\)) in the OB limit of very flat background density profiles  
\(\neref/n_G\approx 1\), \(L_n^{-1} \partial_x \phi \ll 1 \)
and additionally \(\alpha \gg \rhoN^2 k_y/L_n \) (cf.~\cite{camargo98}).
%Scalar measures for non-normality
This is best shown by characterizing the departure from normality by the 
the condition number 
\begin{eqnarray}
    \kappa := \norm{ \vec{V} } \norm{  \vec{V}^{-1}} \geq 1 ,
\end{eqnarray}
of the matrix of eigenvectors \(\vec{V}\) of \(\vec{A}\). Here, the matrix norm is induced by the free energy vector norm \(\norm{\vec{v}}:=\sqrt{ \vec{v}^\dagger \vec{M} \vec{v} }\) with~\cite{camargo98}
\begin{eqnarray}
 \vec{M}:=\frac{1}{2}
 \left(
 \begin{array}{c c}
\vec{I} & \vec{0} \\
\vec{0} &  \rhoN^2\left(- \vec{D}_{xx} + k_y^2 \vec{I}\right) \\
\end{array}\right).
\end{eqnarray}
This normalises the condition number \(\kappa\) to unity for a normal system.
In Fig.~\ref{fig:nonnormality} we plot the condition number \(\kappa\)
as a function of adiabaticity \(\alpha\) and background density length \(L_n\), respectively.  
In contrast to the OB case, we universally obtain non-normality (\(\kappa>1\)) for the NOB case for steep background density gradients (\(L_n=16 \rhoN\)). 
In this regime the NOB condition number \(\kappa\) is at least a magnitude higher than its OB equivalent and approaches extremely large values for adiabaticities below \(\alpha\leq0.01\). For a fixed adiabaticity \(\alpha=0.005\) the NOB dynamics are 
normal in the OB limit (\(L_n \geq 10^4\)), but are strongly non-normal for steep background density gradients. 
This behavior of the condition number \(\kappa\) suggests much larger non-modal effects for the NOB model than for the OB model.
However, the condition number \(\kappa\)  does not give insight into how the departure from normality affects the linear dynamics.
Thus, we analyze in the following the modal and non-modal behavior of Eqs.~\eqref{eq:lindiscrete}. 
%Normal mode analysis
\subsection{Modal analysis}\label{sec:modalanalysis}
%eigenvalues
First, we address the eigenvalues \(\omega(\vec{A})\) and the radial eigenfunctions.
The numerically calculated growth rate \(\gamma(\vec{A}):=Re{(\omega(\vec{A}))}\) and real frequency \(\omega_R(\vec{A}):=-Im{(\omega(\vec{A}))}\) of the NOB case are compared to the maximum of its OB counterpart. 
In the OB limit the dispersion relation can be simply derived analytically~\cite{camargo95, numata07}
\begin{eqnarray}\label{eq:disperion}
0 =  \frac{\omega_{OB}^2}{\OmegaciN^2}   + \frac{\omega_{OB}}{\OmegaciN}  \mathcal{B} +i \frac{\omega_*}{\OmegaciN}   \mathcal{B},
\end{eqnarray}
where we defined  \(\mathcal{B}:= \alpha \left(1+k_\perp^2 \rhoN^2 \right)/(k_\perp^2 \rhoN^2)\), the drift frequency 
\(\omega_* := \OmegaciN k_y \rhoN^2/\left[L_n\left(1+k_\perp^2\rhoN^2\right)\right]\) and
the perpendicular wave number \(k_\perp := \sqrt{ k_x^2 + k_y^2}\).
Consequently, the real frequency \(\omega_{R,OB}\) and growth rate \(\gamma_{OB}\) are derived to
\begin{subequations} 
\begin{eqnarray}
\label{eq:frequency}
\omega_{R,OB} (\vec{k},\alpha,L_n) &:=&  \frac{ \OmegaciN}{2 }  \sqrt{|z|}  \cos{(\theta/2)}  \\
\label{eq:growthrate}
\gamma_{OB} (\vec{k},\alpha,L_n)&:=&   \frac{ \OmegaciN}{2} \left[ -  \mathcal{B}
                   + \sqrt{|z|}  \sin{(\theta/2)}   \right]
\end{eqnarray}
\end{subequations} 
with real part \(Re(z):= - \mathcal{B}^2\), imaginary part \(Im(z) := 4\mathcal{B} \omega_*/\OmegaciN \) and argument \(\theta:=\arg(z)\)
of the complex number \(z\).

\begin{figure*}[ht]
\centering
\includegraphics[width= 0.31\textwidth]{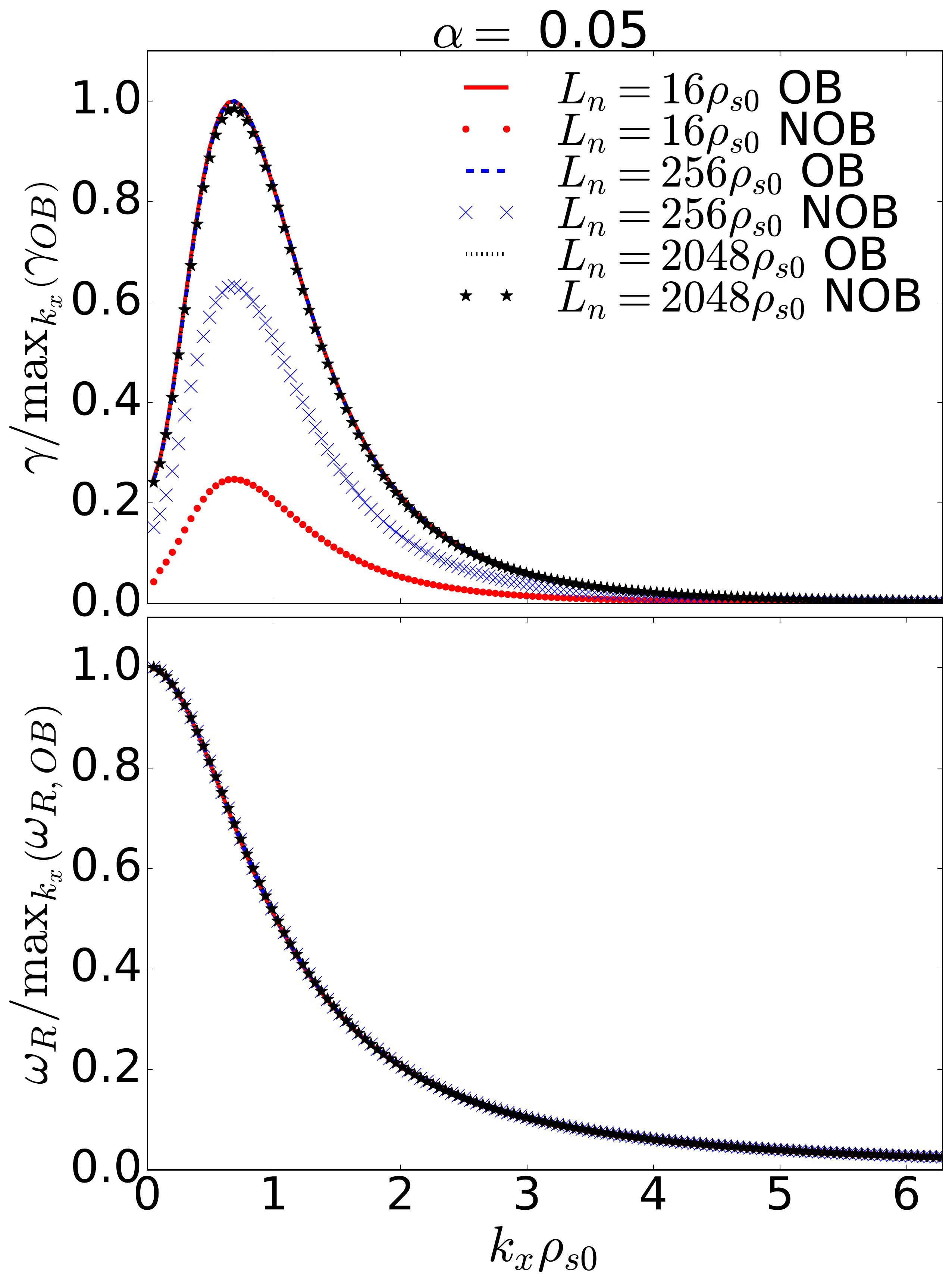}
\includegraphics[width= 0.31\textwidth]{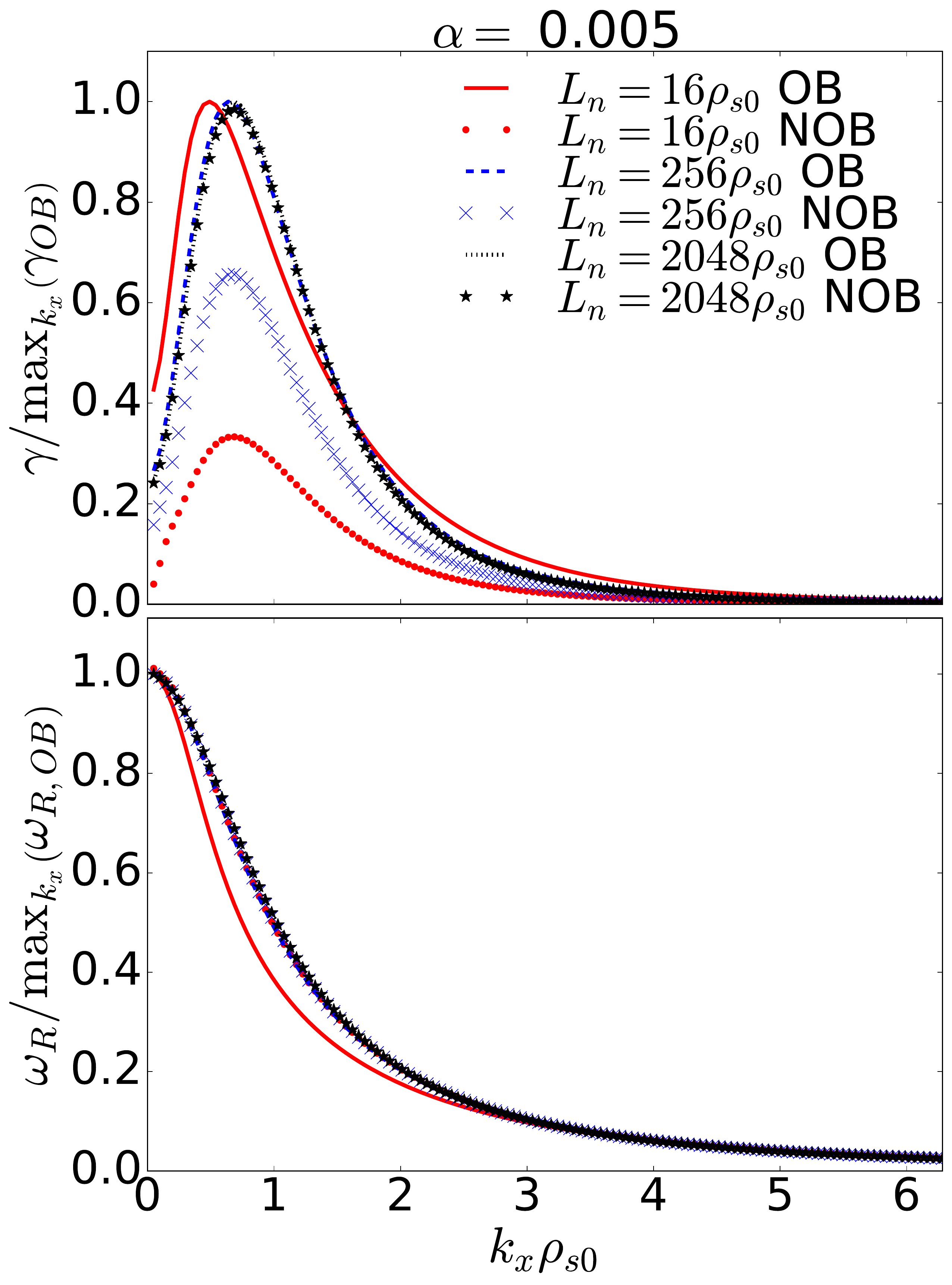}
\includegraphics[width= 0.31\textwidth]{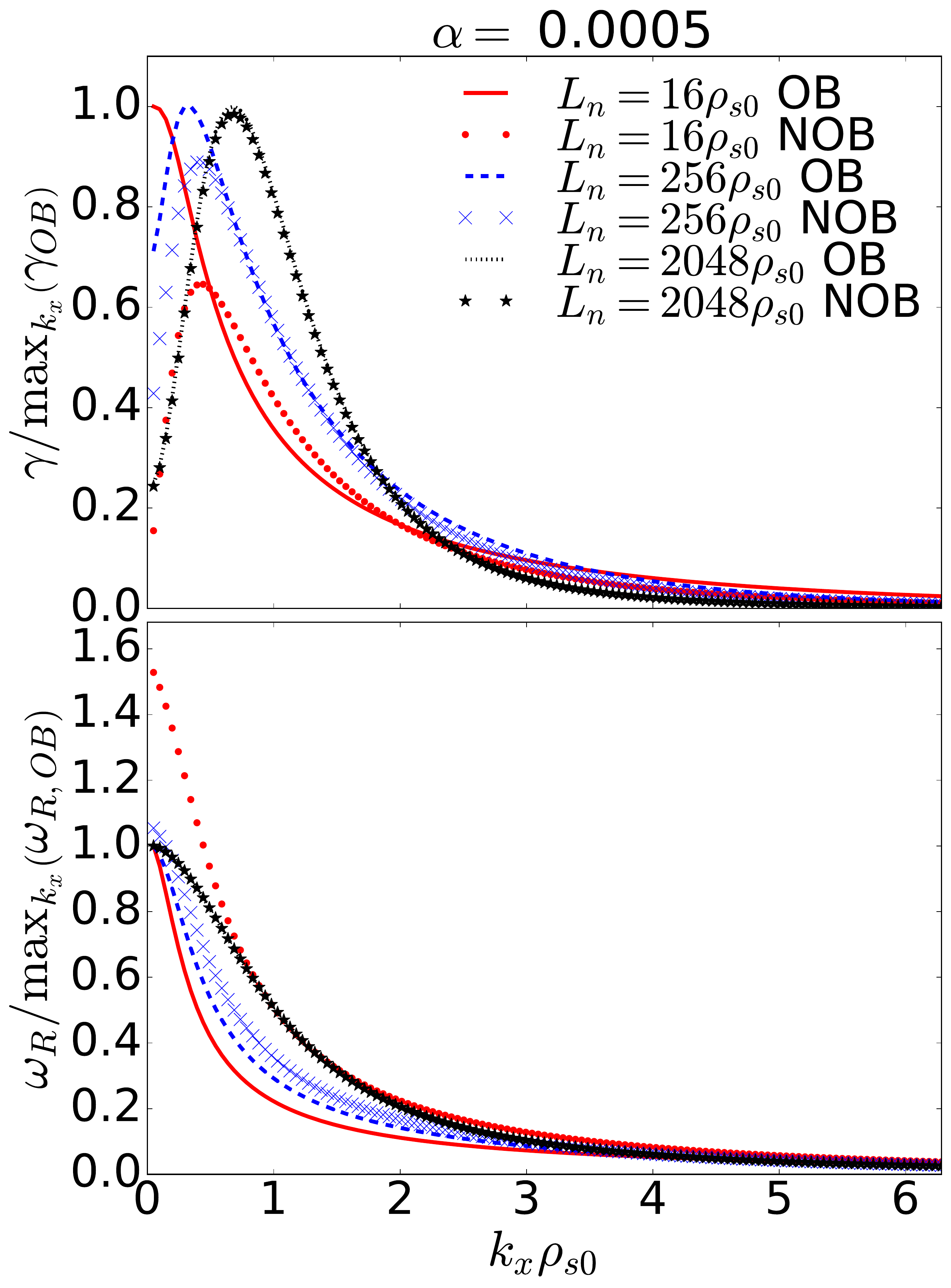}
\caption{The normalised  growth rate \(\gamma/\max_{k_{x}}(\gamma_{OB})\) (top) and real frequency \(\omega_R/\max_{k_{x}}(\omega_{R,OB})\) (bottom) as a function of radial wave-number \(k_x\) 
for  \(k_y = 4 \pi /L_y\) and \(L_x = L_y = 64 \rhoN\) is depicted for various adiabaticity parameters \(\alpha = \left\{0.05, 0.005, 0.0005\right\}\) (left, center, right) and background density gradient lengths \(L_n\). Substantial differences between the NOB and OB growth rates occur for steep background density gradients (\(L_n = 16 \rhoN\)).
}
\label{fig:eigenmodevskx}
\end{figure*}

In Fig.~\ref{fig:eigenmodevskx} we show the normalised growth rate \(\gamma/\max_{k_{x}}(\gamma_{OB})\) and real frequency \(\omega_R/\max_{k_{x}}(\omega_{R,OB})\) as a function of the radial wave number \(k_x\) for various background density gradient lengths \(L_n\) and adiabaticities \(\alpha\). 
Here, the NOB growth rates \(\gamma\) and real frequencies \(\omega_R\) exhibit significant deviations from the
OB limit in particular for steep background gradients and for a range of typical adiabaticity parameters.
In particular, the magnitude of the fastest growing mode differs by up to roughly a factor five and the radial wave-number of the fastest growing mode is also different for certain parameters.
However, the NOB eigenvalues resemble the OB limit for very flat background gradients. 

\begin{figure}[ht]
\centering
 \includegraphics[width= 0.475\textwidth]{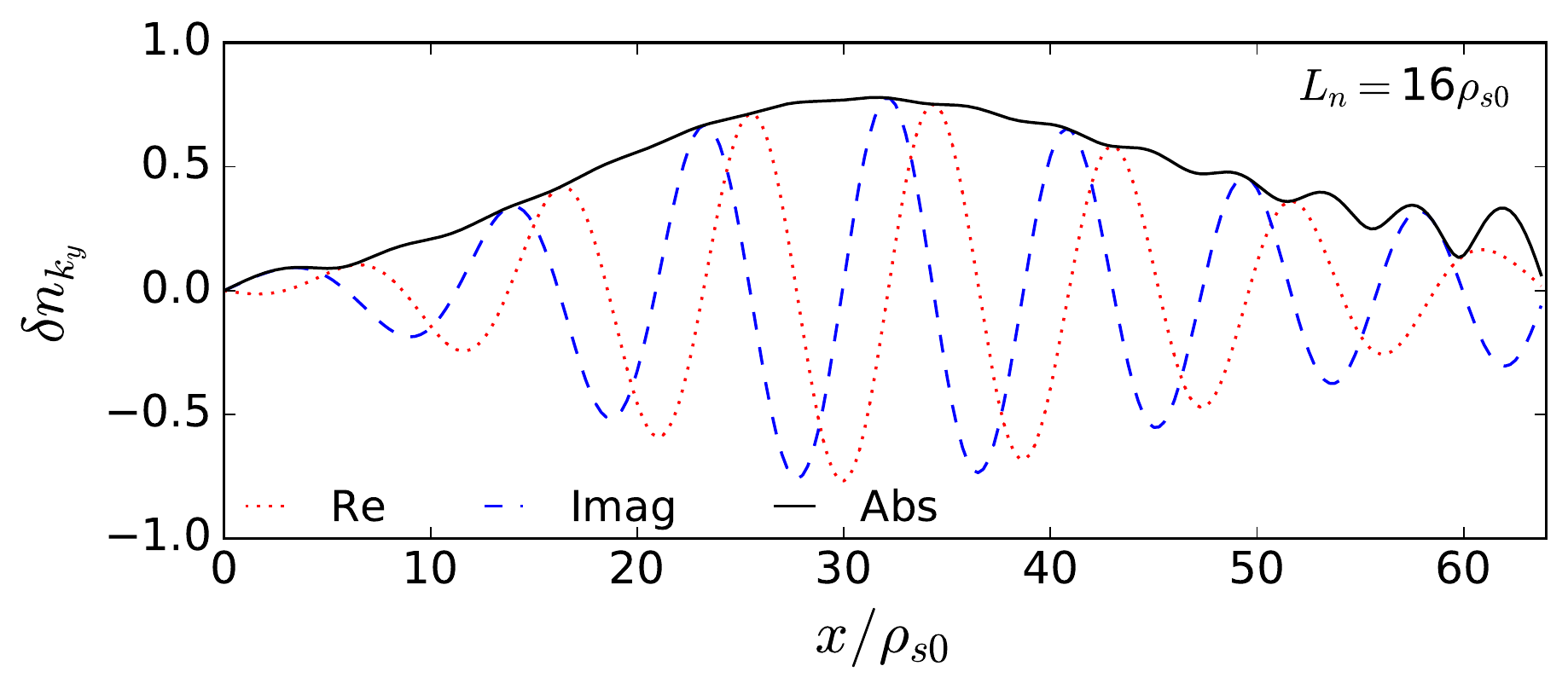}
 \includegraphics[width= 0.475\textwidth]{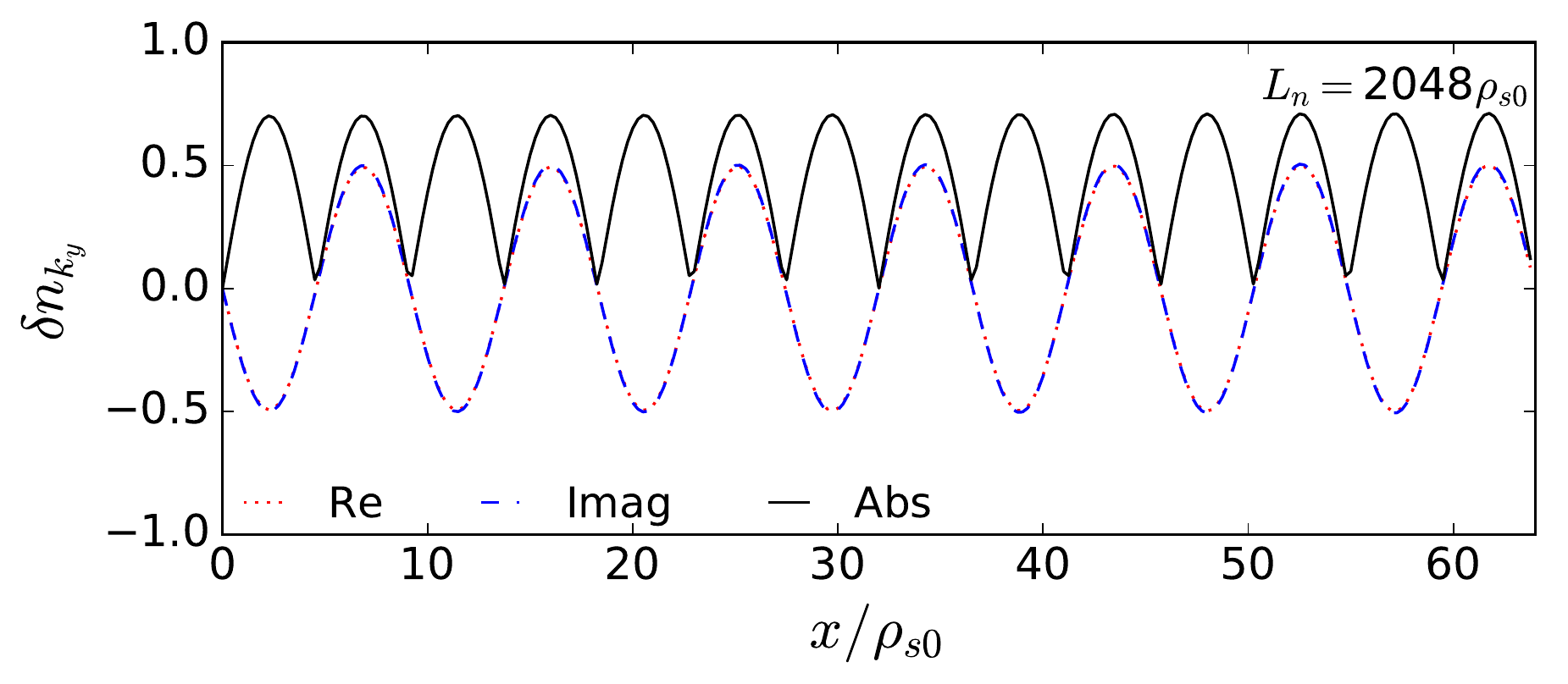}
\caption{The eigenfunctions of the relative density fluctuation \(\flucRrel{\ne}\) for the fastest growing mode for \(k_y = 4 \pi /L_y\), \(L_x = L_y = 64 \rhoN\) and \(\alpha = 0.005\). The background density gradient length is
\(L_n = \left\{16,  2048\right\} \rhoN\) (top,  bottom).
}
\label{fig:eigenfunctionne}
\end{figure}

%eigenfunctions 
The radial eigenfunctions for the relative density fluctuation \(\flucRrel{\ne}_{k_y}\) of the fastest growing \(k_x\)  mode are depicted in Fig.~\ref{fig:eigenfunctionne} for two different density gradient lengths \(L_n\). 
Remarkably, for steep background density gradients these eigenfunctions are spatially localised and do no longer coincide with the ordinary sine like eigenfunction of the OB model. 
Moreover, the phase shift between the real and imaginary parts of the eigenfunctions leads to shearing in the x-y plane as we illustrate in section~\ref{sec:nonmodalanalysis}.
Again, for flat background density profiles the eigenfunctions transition into the OB approximated equivalent.

%Non-modal analysis
\subsection{Non-modal analysis}\label{sec:nonmodalanalysis}
We now face the question how non-modal effects manifest in numerical simulations of the nonlinear full-F ordinary HW model, given by Eqs.~\eqref{eq:fullFdtne},~\eqref{eq:fullFdtNi} and~\eqref{eq:ffpolcoldlwl}.
In particular, we study if initial or transient dynamics are pronounced during the linear phase and if these non-modal effects survive into the nonlinear regime.

%short review of non-modal estimates and growth/decay
It is well known that for a normal matrix the time evolution of the norm of the linear Eq.~\eqref{eq:lindiscrete}
\begin{eqnarray} \label{eq:normbehavior}
 \frac{\norm{ \vec{v}(t)}}{\norm{ \vec{v}(0)} }\leq \norm{ e^{\vec{A} t}},
\end{eqnarray}
is bounded by the spectral abscissa
\begin{eqnarray} \label{eq:specabs}
 \beta(\vec{A}) := 
\max_\vec{k}\left\{\gamma(\vec{A}) \right\},
\end{eqnarray}
for \(t \geq 0\) since the matrix exponential reduces to \( \norm{  e^{ \vec{A} t}} = e^{\beta(\vec{A})  t}\).
However, for a non-normal matrix the maximum growth 
estimate of the modal analysis of section~\ref{sec:modalanalysis}, 
determined by the spectral abscissa \(\beta\),  only holds for \(t \rightarrow \infty\) and the matrix exponential 
reduces to a loose upper bound  \( \norm{  e^{ \vec{A} t}} = \kappa(\vec{V}) e^{\beta(\vec{A})  t}\) for \(t \geq 0\).
As a consequence, a non-normal system may exhibit pronounced initial or transient phenomena, 
for which also estimates and bounds exist. 
In particular,  the numerical abscissa 
\(\eta(\vec{A}) := \max_\vec{k}\left\{\omega\left[(\vec{A}^\dagger+\vec{A})/2\right] \right\} \)\footnote{Note that the spectral and numerical abscissa coincide \(\beta(\vec{A}) = \eta(\vec{A}) \) for a normal matrix (\(\vec{A} \vec{A}^\dagger = \vec{A}^\dagger \vec{A}\)).} represents an upper bound for \(t=0\) and 
the so called \(\epsilon\)-pseudospectral abscissa \(\alpha_\epsilon(\vec{A})\) yields estimates for transient phenomena~\cite{trefethen05}.
Although, the latter two approaches are useful to detect or quantify non-modal effects, we do not make use of them in the following discussion. Instead, we present a direct numerical approach to the initial value problem. 

%Numerical implementation, initialization 
The numerical implementation of the latter full-F gyro-fluid model utilises the open source library \textsc{Feltor}~\cite{feltor41}.
This initial value code relies on a discontinuous Galerkin discretization, which is also used for verification of
the herein presented Galerkin approach for the modal analysis of section~\ref{sec:modalanalysis}. 
We limit our study to a single exemplary initial condition, but note that in general the initial condition can be optimised 
to produce maximum growth at small, intermediate or large time~\cite{camargo98}.
The chosen initial conditions mimics a random perturbation \( \flucRrel{\ne}(\vec{x},0)=\flucRrel{\Ni}(\vec{x},0)= a f_{bath}(\vec{x})\) of amplitude \(a\) with vanishing electric potential \(\phi(\vec{x},0)=0\).
% \footnote{The explicit form of  \(f_{bath}(\vec{x})\) can be found in the documentation of the \textsc{Feltor} library.}.

\begin{figure}[htpb]
\centering
\includegraphics[width= 0.475\textwidth]{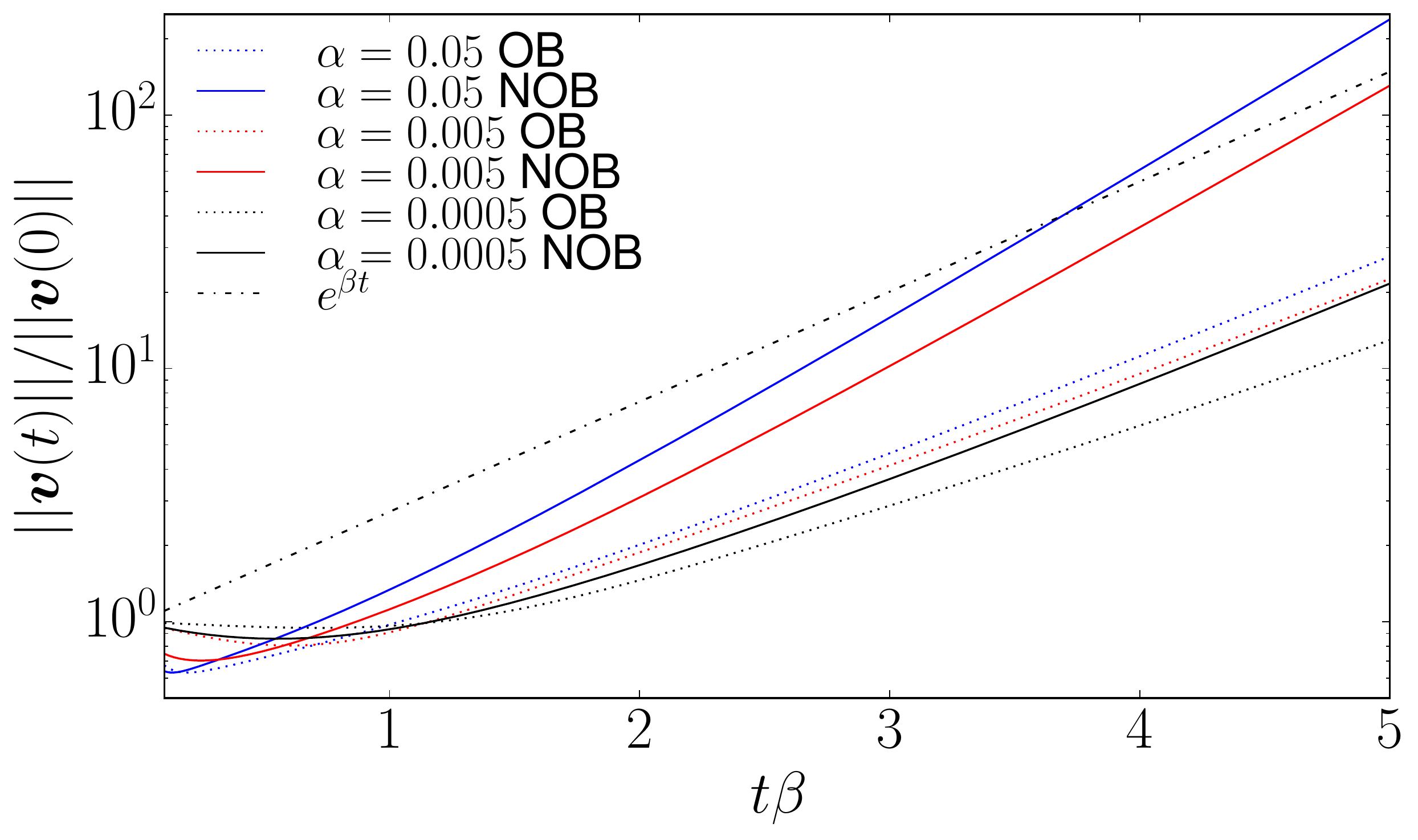}
\caption{The normalised energy norm as a function of normalised time is plotted for different adiabaticities \(\alpha = \left\{0.05,0.005,0.0005\right\} \) in the steep gradient regime \(L_n = 16 \rhoN\). The initial amplitude is \(a=10^{-5}\).}
\label{fig:norm_bath}
\end{figure}

%non-modal effects in the norm behavior
In Fig.~\ref{fig:norm_bath} we show the temporal behavior of the square root of the normalised free energy norm 
\( \norm{  \vec{v}(t)}/\norm{  \vec{v}(0)}  \) for various adiabaticities \(\alpha\) in the steep gradient regime (\(L_n = 16 \rhoN\)). 
Here, the random bath initial condition with the small amplitude \(a=10^{-5}\) limits us to linear effects only.
Interestingly, two clear footprints of non-modal behavior emerge in the linear dynamics. 
First, initial decay of the square root of the normalised free energy norm appears for both the OB and NOB case, despite the fact that all eigenvalues are unstable (cf. Fig.~\ref{fig:eigenmodevskx}). 
Secondly, the transient exponential growth at later times either surpasses (NOB) or falls below (OB and NOB) the spectral abscissa \(\beta\).
Both of these effects are due to the shrinking of non-orthogonal eigenvectors, which is intrinsic to non-normal systems.
We refer the interested reader to~\cite{schmid07} for an illustrative sketch of this phenomenon.
The observed initial decay of the collisional DW instability is similar to that of the collisionless DW instability~\cite{landreman15b}. However, for the latter instability transient amplification can trigger subscritical turbulence in the absence of linear instability. 

\begin{figure}[ht]
\centering
\includegraphics[width= 0.49\textwidth]{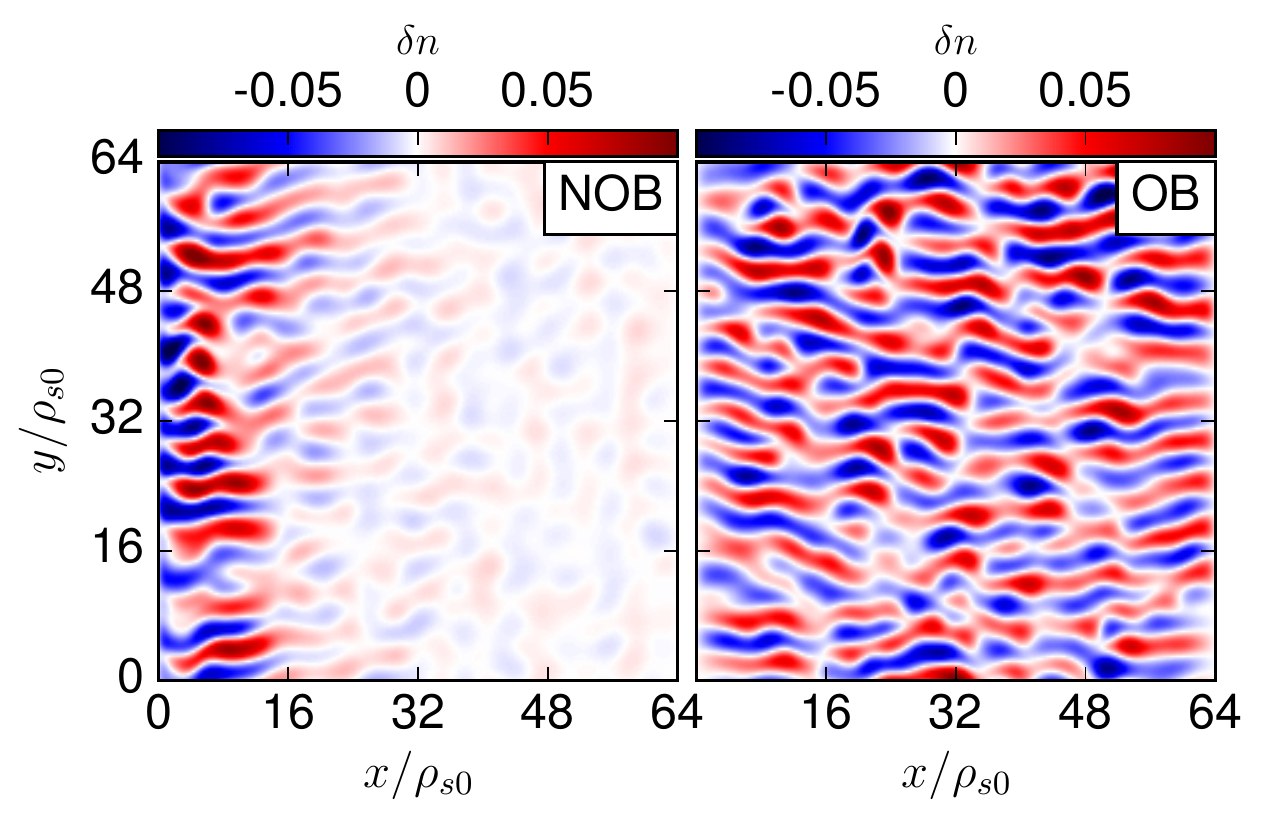}
\caption{The spatial pattern of the relative density fluctuation are shown during the linear phase at 
time \(t=750/\OmegaciN\) (NOB) and \(t=500/\OmegaciN\) (OB).
Localised growth and shearing of the initial relative density perturbation appear in the NOB approximated model. 
The background density gradient length and the adiabaticity is \(L_n = 16 \rhoN\) and \(\alpha=0.005\), respectively. }
\label{fig:linearstructure}
\end{figure}

%non-modal footprint in linear simulations
During this linear growth phase non-modal features may appear in the spatial structure of the relative density fluctuation \(\flucRrel{\ne}\). 
This is depicted in Fig.~\ref{fig:linearstructure}  for the turbulent bath initial condition with \(a=0.01\) and for a steep background density profile (\(L_n = 16 \rhoN\)) and typical adiabaticity (\(\alpha=0.005\)). 
In contrast to the OB approximated model, we observe sheared and localised growth of the initial perturbation in the steep background density regime. 
This is in qualitativ agreement with the previously reported NOB shearing effect of DWs of~\cite{held15}.
The radial location of the strongest growth coincides approximately with the maximum of the absolute background density gradient \(| \partial_x n_G |\). These NOB effects are again reasoned in the strong non-orthogonality of the eigenvector Matrix \(\vec{V}\), which is pronounced for
a system with strong non-normality and by implication high condition number \(\kappa\) (cf. Fig.~\ref{fig:nonnormality})
% \footnote{Non-orthogonality can be measued by the deformation angle \(\theta\), which can be defined as \(\theta:=2 \arccot{(\kappa)}\)~\cite{trefethen05}.}.

\begin{figure}[ht]
\centering
\includegraphics[width= 0.49\textwidth]{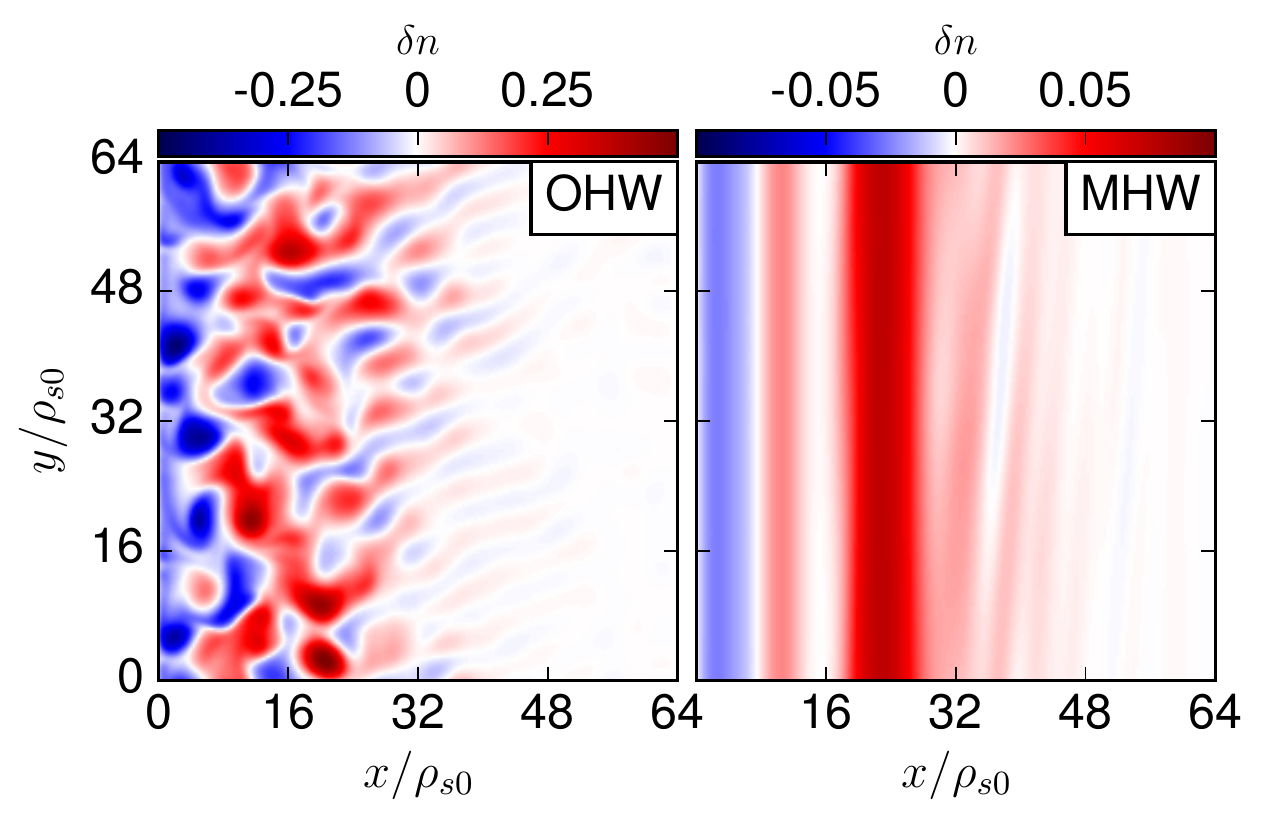}
\caption{The spatial pattern of the relative density fluctuation \( \flucRrel{\ne}\) is shown during the nonlinear phase without (OHW)  and with zonal flows (MHW), at time \(t=4500/\OmegaciN\) and \(t=9500/\OmegaciN\), respectively.
Radial peaking of the turbulence intensity in the nonlinear regime  proves to be a NOB effect. 
The background density gradient length and the adiabaticity is \(L_n = 16 \rhoN\) and \(\alpha=0.005\), respectively. }
\label{fig:nonlinearstructure}
\end{figure}

%Non-modal footprint in nonlinear simulations
Finally, we study how non-modality affects the turbulence intensity in collisional DW turbulence without and with zonal flows. Here, the underlying models are the NOB-extended OHW (equations~\eqref{eq:fullFdtne},~\eqref{eq:fullFdtNi} and~\eqref{eq:ffpolcoldlwl}) and modified HW (MHW) model~\cite{held18}, respectively.
In Fig.~\ref{fig:nonlinearstructure} we show the spatial structure of the relative density fluctuation \(\flucRrel{\ne}\) for both models and the latter initial condition but during the nonlinear phase.  
Without zonal flows (OHW) we observe a radial peaking of the  maximum of the relative density fluctuation amplitude. 
Analogously, with zonal flows (MHW) a similar radial peaking appears for the zonal flow amplitude. 
Note that the radial localization of the turbulence intensity continues to exist at late turbulence saturation times if the radial particle transport is weak. This occurs for large adiabaticity (small collisionality). 
For both model cases, this constitutes a clear non-modal footprint in saturated collisional DW turbulence and shows that indeed these non-modal effects can survive into the nonlinear regime. 
As opposed to this, the radial peaking of the turbulence or zonal flow intensity is again absent in the OB approximated OHW or MHW model (cf.~\cite{camargo95,numata07,kendl18}).
%Discussion and conclusion
\section{Disussion and conclusion}
%summary
We studied the collisional DW instability in a straight and unsheared magnetic field within a full-F gyro-fluid model, which relaxes the OB approximation. In the regime of steep background density gradients both the eigenvalues and eigenfunctions fundamentally deviated from former OB approximated investigations. In particular, our modal analysis demonstrated NOB corrections by factors of order one to the eigenvalues, highly non-orthogonal eigenvectors and spatially localised eigenfunctions for typical plasma parameters.
Our non-modal analysis revealed initial damping and transient non-modal growth of the free energy of an initially unstable random perturbation. 
Remarkably, this linear growth is radially localised and sheared. 
It was numerically shown that this NOB signature subsists into the nonlinear regime, where radially localised turbulence or zonal flow amplitudes emerge. 

%%Discussion
The herein presented results emphasise the need for NOB approximated models to consistently capture the (linear) dynamics of the collisional DW instability for 
large density inhomogeneities. For instance, this may prove necessary for the accurate calculation of transport levels in high-confinement tokamak plasmas within quasilinear gyro-kinetic or gyro-fluid models (e.g.:~\cite{staebler05,bourdelle07}).
Finally, we conclude that 
the study of linear effects, that is solely based on a modal approach, may give misleading predictions since this approach overlooks non-modal features like initial and transient phenomena.

\section{Acknowledgements}
This work was supported by the Austrian Science Fund (FWF) Y398. The computational results presented have been achieved using the Vienna Scientific Cluster (VSC) and 
the EUROfusion High Performance Computer (Marconi-Fusion).
% \section*{References}
\bibliography{meanflows_aip.bib}

%merlin.mbs aipnum4-1.bst 2010-07-25 4.21a (PWD, AO, DPC) hacked
%Control: key (0)
%Control: author (8) initials jnrlst
%Control: editor formatted (1) identically to author
%Control: production of article title (-1) disabled
%Control: page (0) single
%Control: year (1) truncated
%Control: production of eprint (0) enabled
\begin{thebibliography}{53}%
\makeatletter
\providecommand \@ifxundefined [1]{%
 \@ifx{#1\undefined}
}%
\providecommand \@ifnum [1]{%
 \ifnum #1\expandafter \@firstoftwo
 \else \expandafter \@secondoftwo
 \fi
}%
\providecommand \@ifx [1]{%
 \ifx #1\expandafter \@firstoftwo
 \else \expandafter \@secondoftwo
 \fi
}%
\providecommand \natexlab [1]{#1}%
\providecommand \enquote  [1]{``#1''}%
\providecommand \bibnamefont  [1]{#1}%
\providecommand \bibfnamefont [1]{#1}%
\providecommand \citenamefont [1]{#1}%
\providecommand \href@noop [0]{\@secondoftwo}%
\providecommand \href [0]{\begingroup \@sanitize@url \@href}%
\providecommand \@href[1]{\@@startlink{#1}\@@href}%
\providecommand \@@href[1]{\endgroup#1\@@endlink}%
\providecommand \@sanitize@url [0]{\catcode `\\12\catcode `\$12\catcode
  `\&12\catcode `\#12\catcode `\^12\catcode `\_12\catcode `\%12\relax}%
\providecommand \@@startlink[1]{}%
\providecommand \@@endlink[0]{}%
\providecommand \url  [0]{\begingroup\@sanitize@url \@url }%
\providecommand \@url [1]{\endgroup\@href {#1}{\urlprefix }}%
\providecommand \urlprefix  [0]{URL }%
\providecommand \Eprint [0]{\href }%
\providecommand \doibase [0]{http://dx.doi.org/}%
\providecommand \selectlanguage [0]{\@gobble}%
\providecommand \bibinfo  [0]{\@secondoftwo}%
\providecommand \bibfield  [0]{\@secondoftwo}%
\providecommand \translation [1]{[#1]}%
\providecommand \BibitemOpen [0]{}%
\providecommand \bibitemStop [0]{}%
\providecommand \bibitemNoStop [0]{.\EOS\space}%
\providecommand \EOS [0]{\spacefactor3000\relax}%
\providecommand \BibitemShut  [1]{\csname bibitem#1\endcsname}%
\let\auto@bib@innerbib\@empty
%</preamble>
\bibitem [{\citenamefont {{Rudakov}}\ and\ \citenamefont
  {{Sagdeev}}(1961)}]{rudakov61}%
  \BibitemOpen
  \bibfield  {author} {\bibinfo {author} {\bibfnamefont {L.~I.}\ \bibnamefont
  {{Rudakov}}}\ and\ \bibinfo {author} {\bibfnamefont {R.~Z.}\ \bibnamefont
  {{Sagdeev}}},\ }\href@noop {} {\bibfield  {journal} {\bibinfo  {journal}
  {Soviet Physics Doklady}\ }\textbf {\bibinfo {volume} {6}},\ \bibinfo {pages}
  {415} (\bibinfo {year} {1961})}\BibitemShut {NoStop}%
\bibitem [{\citenamefont {Chen}(1964{\natexlab{a}})}]{chen63}%
  \BibitemOpen
  \bibfield  {author} {\bibinfo {author} {\bibfnamefont {F.~F.}\ \bibnamefont
  {Chen}},\ }\href@noop {} {\bibfield  {journal} {\bibinfo  {journal} {Proc. of
  the Sixth Conf. on Ioniz. Phenomena in Gases}\ }\textbf {\bibinfo {volume}
  {2}},\ \bibinfo {pages} {435} (\bibinfo {year}
  {1964}{\natexlab{a}})}\BibitemShut {NoStop}%
\bibitem [{\citenamefont {Chen}(1964{\natexlab{b}})}]{chen64}%
  \BibitemOpen
  \bibfield  {author} {\bibinfo {author} {\bibfnamefont {F.~F.}\ \bibnamefont
  {Chen}},\ }\href {\doibase 10.1063/1.1711341} {\bibfield  {journal} {\bibinfo
   {journal} {The Physics of Fluids}\ }\textbf {\bibinfo {volume} {7}},\
  \bibinfo {pages} {949} (\bibinfo {year} {1964}{\natexlab{b}})}\BibitemShut
  {NoStop}%
\bibitem [{\citenamefont {Moiseev}\ and\ \citenamefont
  {Sagdeev}(1963)}]{moiseev63}%
  \BibitemOpen
  \bibfield  {author} {\bibinfo {author} {\bibfnamefont {S.~S.}\ \bibnamefont
  {Moiseev}}\ and\ \bibinfo {author} {\bibfnamefont {R.~Z.}\ \bibnamefont
  {Sagdeev}},\ }\href@noop {} {\bibfield  {journal} {\bibinfo  {journal}
  {Soviet Physics JETP-USSR}\ }\textbf {\bibinfo {volume} {17}},\ \bibinfo
  {pages} {515} (\bibinfo {year} {1963})}\BibitemShut {NoStop}%
\bibitem [{\citenamefont {Galeev}, \citenamefont {Oraevskii},\ and\
  \citenamefont {Sagdeev}(1963)}]{galeev63}%
  \BibitemOpen
  \bibfield  {author} {\bibinfo {author} {\bibfnamefont {A.}~\bibnamefont
  {Galeev}}, \bibinfo {author} {\bibfnamefont {V.}~\bibnamefont {Oraevskii}}, \
  and\ \bibinfo {author} {\bibfnamefont {R.}~\bibnamefont {Sagdeev}},\
  }\href@noop {} {\bibfield  {journal} {\bibinfo  {journal} {Zh. Eksperim. i
  Teor. Fiz.}\ }\textbf {\bibinfo {volume} {44}} (\bibinfo {year}
  {1963})}\BibitemShut {NoStop}%
\bibitem [{\citenamefont {Krall}\ and\ \citenamefont
  {Rosenbluth}(1965)}]{krall65}%
  \BibitemOpen
  \bibfield  {author} {\bibinfo {author} {\bibfnamefont {N.~A.}\ \bibnamefont
  {Krall}}\ and\ \bibinfo {author} {\bibfnamefont {M.~N.}\ \bibnamefont
  {Rosenbluth}},\ }\href {\doibase 10.1063/1.1761444} {\bibfield  {journal}
  {\bibinfo  {journal} {The Physics of Fluids}\ }\textbf {\bibinfo {volume}
  {8}},\ \bibinfo {pages} {1488} (\bibinfo {year} {1965})}\BibitemShut
  {NoStop}%
\bibitem [{\citenamefont {Kadomtsev}\ and\ \citenamefont
  {Pogutse}(1970)}]{kadomtsev70}%
  \BibitemOpen
  \bibfield  {author} {\bibinfo {author} {\bibfnamefont {B.}~\bibnamefont
  {Kadomtsev}}\ and\ \bibinfo {author} {\bibfnamefont {O.}~\bibnamefont
  {Pogutse}},\ }in\ \href@noop {} {\emph {\bibinfo {booktitle} {Reviews of
  plasma physics}}}\ (\bibinfo  {publisher} {Springer},\ \bibinfo {year}
  {1970})\ pp.\ \bibinfo {pages} {249--400}\BibitemShut {NoStop}%
\bibitem [{\citenamefont {Mikhailovskii}(1974)}]{mikhailovskii74}%
  \BibitemOpen
  \bibfield  {author} {\bibinfo {author} {\bibfnamefont {A.~B.}\ \bibnamefont
  {Mikhailovskii}},\ }\href@noop {} {\emph {\bibinfo {title} {Theory of Plasma
  Instabilities, Vol. 2: Instabilities of an Inhomogeneous Plasma}}}\ (\bibinfo
   {publisher} {Springer US},\ \bibinfo {year} {1974})\BibitemShut {NoStop}%
\bibitem [{\citenamefont {Tang}(1978)}]{tang78}%
  \BibitemOpen
  \bibfield  {author} {\bibinfo {author} {\bibfnamefont {W.}~\bibnamefont
  {Tang}},\ }\href {\doibase 10.1088/0029-5515/18/8/006} {\bibfield  {journal}
  {\bibinfo  {journal} {Nuclear Fusion}\ }\textbf {\bibinfo {volume} {18}},\
  \bibinfo {pages} {1089} (\bibinfo {year} {1978})}\BibitemShut {NoStop}%
\bibitem [{\citenamefont {Horton}(1999)}]{horton99}%
  \BibitemOpen
  \bibfield  {author} {\bibinfo {author} {\bibfnamefont {W.}~\bibnamefont
  {Horton}},\ }\href {\doibase 10.1103/RevModPhys.71.735} {\bibfield  {journal}
  {\bibinfo  {journal} {Rev. Mod. Phys.}\ }\textbf {\bibinfo {volume} {71}},\
  \bibinfo {pages} {735} (\bibinfo {year} {1999})}\BibitemShut {NoStop}%
\bibitem [{\citenamefont {Bellan}()}]{bellan06}%
  \BibitemOpen
  \bibfield  {author} {\bibinfo {author} {\bibfnamefont {P.~M.}\ \bibnamefont
  {Bellan}},\ }\href {\doibase 10.1017/CBO9780511807183} {\emph {\bibinfo
  {title} {Fundamentals of Plasma Physics}}}\ (\bibinfo  {publisher} {Cambridge
  University Press})\BibitemShut {NoStop}%
\bibitem [{\citenamefont {Guzdar}\ \emph {et~al.}(1978)\citenamefont {Guzdar},
  \citenamefont {Chen}, \citenamefont {Kaw},\ and\ \citenamefont
  {Oberman}}]{guzdar78}%
  \BibitemOpen
  \bibfield  {author} {\bibinfo {author} {\bibfnamefont {P.~N.}\ \bibnamefont
  {Guzdar}}, \bibinfo {author} {\bibfnamefont {L.}~\bibnamefont {Chen}},
  \bibinfo {author} {\bibfnamefont {P.~K.}\ \bibnamefont {Kaw}}, \ and\
  \bibinfo {author} {\bibfnamefont {C.}~\bibnamefont {Oberman}},\ }\href
  {\doibase 10.1103/PhysRevLett.40.1566} {\bibfield  {journal} {\bibinfo
  {journal} {Phys. Rev. Lett.}\ }\textbf {\bibinfo {volume} {40}},\ \bibinfo
  {pages} {1566} (\bibinfo {year} {1978})}\BibitemShut {NoStop}%
\bibitem [{\citenamefont {Chen}\ \emph {et~al.}(1979)\citenamefont {Chen},
  \citenamefont {Guzdar}, \citenamefont {Hsu}, \citenamefont {Kaw},
  \citenamefont {Oberman},\ and\ \citenamefont {White}}]{chen79}%
  \BibitemOpen
  \bibfield  {author} {\bibinfo {author} {\bibfnamefont {L.}~\bibnamefont
  {Chen}}, \bibinfo {author} {\bibfnamefont {P.}~\bibnamefont {Guzdar}},
  \bibinfo {author} {\bibfnamefont {J.}~\bibnamefont {Hsu}}, \bibinfo {author}
  {\bibfnamefont {P.}~\bibnamefont {Kaw}}, \bibinfo {author} {\bibfnamefont
  {C.}~\bibnamefont {Oberman}}, \ and\ \bibinfo {author} {\bibfnamefont
  {R.}~\bibnamefont {White}},\ }\href {\doibase 10.1088/0029-5515/19/3/009}
  {\bibfield  {journal} {\bibinfo  {journal} {Nuclear Fusion}\ }\textbf
  {\bibinfo {volume} {19}},\ \bibinfo {pages} {373} (\bibinfo {year}
  {1979})}\BibitemShut {NoStop}%
\bibitem [{\citenamefont {Pearlstein}\ and\ \citenamefont
  {Berk}(1969)}]{pearlstein69}%
  \BibitemOpen
  \bibfield  {author} {\bibinfo {author} {\bibfnamefont {L.~D.}\ \bibnamefont
  {Pearlstein}}\ and\ \bibinfo {author} {\bibfnamefont {H.~L.}\ \bibnamefont
  {Berk}},\ }\href {\doibase 10.1103/PhysRevLett.23.220} {\bibfield  {journal}
  {\bibinfo  {journal} {Phys. Rev. Lett.}\ }\textbf {\bibinfo {volume} {23}},\
  \bibinfo {pages} {220} (\bibinfo {year} {1969})}\BibitemShut {NoStop}%
\bibitem [{\citenamefont {Tsang}\ \emph {et~al.}(1978)\citenamefont {Tsang},
  \citenamefont {Catto}, \citenamefont {Whitson},\ and\ \citenamefont
  {Smith}}]{tsang78}%
  \BibitemOpen
  \bibfield  {author} {\bibinfo {author} {\bibfnamefont {K.~T.}\ \bibnamefont
  {Tsang}}, \bibinfo {author} {\bibfnamefont {P.~J.}\ \bibnamefont {Catto}},
  \bibinfo {author} {\bibfnamefont {J.~C.}\ \bibnamefont {Whitson}}, \ and\
  \bibinfo {author} {\bibfnamefont {J.}~\bibnamefont {Smith}},\ }\href
  {\doibase 10.1103/PhysRevLett.40.327} {\bibfield  {journal} {\bibinfo
  {journal} {Phys. Rev. Lett.}\ }\textbf {\bibinfo {volume} {40}},\ \bibinfo
  {pages} {327} (\bibinfo {year} {1978})}\BibitemShut {NoStop}%
\bibitem [{\citenamefont {Ross}\ and\ \citenamefont {Mahajan}(1978)}]{ross78}%
  \BibitemOpen
  \bibfield  {author} {\bibinfo {author} {\bibfnamefont {D.~W.}\ \bibnamefont
  {Ross}}\ and\ \bibinfo {author} {\bibfnamefont {S.~M.}\ \bibnamefont
  {Mahajan}},\ }\href {\doibase 10.1103/PhysRevLett.40.324} {\bibfield
  {journal} {\bibinfo  {journal} {Phys. Rev. Lett.}\ }\textbf {\bibinfo
  {volume} {40}},\ \bibinfo {pages} {324} (\bibinfo {year} {1978})}\BibitemShut
  {NoStop}%
\bibitem [{\citenamefont {Chen}\ \emph {et~al.}(1978)\citenamefont {Chen},
  \citenamefont {Guzdar}, \citenamefont {White}, \citenamefont {Kaw},\ and\
  \citenamefont {Oberman}}]{chen78}%
  \BibitemOpen
  \bibfield  {author} {\bibinfo {author} {\bibfnamefont {L.}~\bibnamefont
  {Chen}}, \bibinfo {author} {\bibfnamefont {P.~N.}\ \bibnamefont {Guzdar}},
  \bibinfo {author} {\bibfnamefont {R.~B.}\ \bibnamefont {White}}, \bibinfo
  {author} {\bibfnamefont {P.~K.}\ \bibnamefont {Kaw}}, \ and\ \bibinfo
  {author} {\bibfnamefont {C.}~\bibnamefont {Oberman}},\ }\href {\doibase
  10.1103/PhysRevLett.41.649} {\bibfield  {journal} {\bibinfo  {journal} {Phys.
  Rev. Lett.}\ }\textbf {\bibinfo {volume} {41}},\ \bibinfo {pages} {649}
  (\bibinfo {year} {1978})}\BibitemShut {NoStop}%
\bibitem [{\citenamefont {Antonsen}(1978)}]{antonsen78}%
  \BibitemOpen
  \bibfield  {author} {\bibinfo {author} {\bibfnamefont {T.~M.}\ \bibnamefont
  {Antonsen}},\ }\href {\doibase 10.1103/PhysRevLett.41.33} {\bibfield
  {journal} {\bibinfo  {journal} {Phys. Rev. Lett.}\ }\textbf {\bibinfo
  {volume} {41}},\ \bibinfo {pages} {33} (\bibinfo {year} {1978})}\BibitemShut
  {NoStop}%
\bibitem [{\citenamefont {Landreman}, \citenamefont {Antonsen},\ and\
  \citenamefont {Dorland}(2015)}]{landreman15}%
  \BibitemOpen
  \bibfield  {author} {\bibinfo {author} {\bibfnamefont {M.}~\bibnamefont
  {Landreman}}, \bibinfo {author} {\bibfnamefont {T.~M.}\ \bibnamefont
  {Antonsen}}, \ and\ \bibinfo {author} {\bibfnamefont {W.}~\bibnamefont
  {Dorland}},\ }\href {\doibase 10.1103/PhysRevLett.114.095003} {\bibfield
  {journal} {\bibinfo  {journal} {Phys. Rev. Lett.}\ }\textbf {\bibinfo
  {volume} {114}},\ \bibinfo {pages} {095003} (\bibinfo {year}
  {2015})}\BibitemShut {NoStop}%
\bibitem [{\citenamefont {Helander}\ and\ \citenamefont
  {Plunk}(2015)}]{helander15}%
  \BibitemOpen
  \bibfield  {author} {\bibinfo {author} {\bibfnamefont {P.}~\bibnamefont
  {Helander}}\ and\ \bibinfo {author} {\bibfnamefont {G.~G.}\ \bibnamefont
  {Plunk}},\ }\href {\doibase 10.1063/1.4932081} {\bibfield  {journal}
  {\bibinfo  {journal} {Physics of Plasmas}\ }\textbf {\bibinfo {volume}
  {22}},\ \bibinfo {pages} {090706} (\bibinfo {year} {2015})}\BibitemShut
  {NoStop}%
\bibitem [{\citenamefont {Camargo}, \citenamefont {Tippett},\ and\
  \citenamefont {Caldas}(1998)}]{camargo98}%
  \BibitemOpen
  \bibfield  {author} {\bibinfo {author} {\bibfnamefont {S.~J.}\ \bibnamefont
  {Camargo}}, \bibinfo {author} {\bibfnamefont {M.~K.}\ \bibnamefont
  {Tippett}}, \ and\ \bibinfo {author} {\bibfnamefont {I.~L.}\ \bibnamefont
  {Caldas}},\ }\href {\doibase 10.1103/PhysRevE.58.3693} {\bibfield  {journal}
  {\bibinfo  {journal} {Phys. Rev. E}\ }\textbf {\bibinfo {volume} {58}},\
  \bibinfo {pages} {3693} (\bibinfo {year} {1998})}\BibitemShut {NoStop}%
\bibitem [{\citenamefont {Mikhailenko}, \citenamefont {Mikhailenko},\ and\
  \citenamefont {Stepanov}(2000)}]{mikhailenko00}%
  \BibitemOpen
  \bibfield  {author} {\bibinfo {author} {\bibfnamefont {V.~S.}\ \bibnamefont
  {Mikhailenko}}, \bibinfo {author} {\bibfnamefont {V.~V.}\ \bibnamefont
  {Mikhailenko}}, \ and\ \bibinfo {author} {\bibfnamefont {K.~N.}\ \bibnamefont
  {Stepanov}},\ }\href {\doibase 10.1063/1.873785} {\bibfield  {journal}
  {\bibinfo  {journal} {Physics of Plasmas}\ }\textbf {\bibinfo {volume} {7}},\
  \bibinfo {pages} {94} (\bibinfo {year} {2000})}\BibitemShut {NoStop}%
\bibitem [{\citenamefont {Friedman}\ and\ \citenamefont
  {Carter}(2015)}]{friedman15}%
  \BibitemOpen
  \bibfield  {author} {\bibinfo {author} {\bibfnamefont {B.}~\bibnamefont
  {Friedman}}\ and\ \bibinfo {author} {\bibfnamefont {T.~A.}\ \bibnamefont
  {Carter}},\ }\href {\doibase 10.1063/1.4905863} {\bibfield  {journal}
  {\bibinfo  {journal} {Physics of Plasmas}\ }\textbf {\bibinfo {volume}
  {22}},\ \bibinfo {pages} {012307} (\bibinfo {year} {2015})}\BibitemShut
  {NoStop}%
\bibitem [{\citenamefont {Scott}(1990)}]{scott90}%
  \BibitemOpen
  \bibfield  {author} {\bibinfo {author} {\bibfnamefont {B.~D.}\ \bibnamefont
  {Scott}},\ }\href {\doibase 10.1103/PhysRevLett.65.3289} {\bibfield
  {journal} {\bibinfo  {journal} {Phys. Rev. Lett.}\ }\textbf {\bibinfo
  {volume} {65}},\ \bibinfo {pages} {3289} (\bibinfo {year}
  {1990})}\BibitemShut {NoStop}%
\bibitem [{\citenamefont {Scott}(1992)}]{scott92}%
  \BibitemOpen
  \bibfield  {author} {\bibinfo {author} {\bibfnamefont {B.~D.}\ \bibnamefont
  {Scott}},\ }\href {\doibase 10.1063/1.860215} {\bibfield  {journal} {\bibinfo
   {journal} {Physics of Fluids B: Plasma Physics}\ }\textbf {\bibinfo {volume}
  {4}},\ \bibinfo {pages} {2468} (\bibinfo {year} {1992})}\BibitemShut
  {NoStop}%
\bibitem [{\citenamefont {Drake}, \citenamefont {Zeiler},\ and\ \citenamefont
  {Biskamp}(1995)}]{drake95}%
  \BibitemOpen
  \bibfield  {author} {\bibinfo {author} {\bibfnamefont {J.~F.}\ \bibnamefont
  {Drake}}, \bibinfo {author} {\bibfnamefont {A.}~\bibnamefont {Zeiler}}, \
  and\ \bibinfo {author} {\bibfnamefont {D.}~\bibnamefont {Biskamp}},\ }\href
  {\doibase 10.1103/PhysRevLett.75.4222} {\bibfield  {journal} {\bibinfo
  {journal} {Phys. Rev. Lett.}\ }\textbf {\bibinfo {volume} {75}},\ \bibinfo
  {pages} {4222} (\bibinfo {year} {1995})}\BibitemShut {NoStop}%
\bibitem [{\citenamefont {Landreman}, \citenamefont {Plunk},\ and\
  \citenamefont {Dorland}(2015)}]{landreman15b}%
  \BibitemOpen
  \bibfield  {author} {\bibinfo {author} {\bibfnamefont {M.}~\bibnamefont
  {Landreman}}, \bibinfo {author} {\bibfnamefont {G.~G.}\ \bibnamefont
  {Plunk}}, \ and\ \bibinfo {author} {\bibfnamefont {W.}~\bibnamefont
  {Dorland}},\ }\href {\doibase 10.1017/S0022377815000495} {\bibfield
  {journal} {\bibinfo  {journal} {Journal of Plasma Physics}\ }\textbf
  {\bibinfo {volume} {81}} (\bibinfo {year} {2015}),\
  10.1017/S0022377815000495}\BibitemShut {NoStop}%
\bibitem [{\citenamefont {Angus}\ and\ \citenamefont
  {Krasheninnikov}(2012)}]{angus12}%
  \BibitemOpen
  \bibfield  {author} {\bibinfo {author} {\bibfnamefont {J.~R.}\ \bibnamefont
  {Angus}}\ and\ \bibinfo {author} {\bibfnamefont {S.~I.}\ \bibnamefont
  {Krasheninnikov}},\ }\href {\doibase 10.1063/1.4714614} {\bibfield  {journal}
  {\bibinfo  {journal} {Physics of Plasmas}\ }\textbf {\bibinfo {volume}
  {19}},\ \bibinfo {pages} {052504} (\bibinfo {year} {2012})}\BibitemShut
  {NoStop}%
\bibitem [{\citenamefont {Jorge}, \citenamefont {Ricci},\ and\ \citenamefont
  {Loureiro}(2018)}]{jorge18}%
  \BibitemOpen
  \bibfield  {author} {\bibinfo {author} {\bibfnamefont {R.}~\bibnamefont
  {Jorge}}, \bibinfo {author} {\bibfnamefont {P.}~\bibnamefont {Ricci}}, \ and\
  \bibinfo {author} {\bibfnamefont {N.~F.}\ \bibnamefont {Loureiro}},\ }\href
  {\doibase 10.1103/PhysRevLett.121.165001} {\bibfield  {journal} {\bibinfo
  {journal} {Phys. Rev. Lett.}\ }\textbf {\bibinfo {volume} {121}},\ \bibinfo
  {pages} {165001} (\bibinfo {year} {2018})}\BibitemShut {NoStop}%
\bibitem [{\citenamefont {Shao}\ \emph {et~al.}(2016)\citenamefont {Shao},
  \citenamefont {Wolfrum}, \citenamefont {Ryter}, \citenamefont {Birkenmeier},
  \citenamefont {Laggner}, \citenamefont {Viezzer}, \citenamefont {Fischer},
  \citenamefont {Willensdorfer}, \citenamefont {Kurzan}, \citenamefont {Lunt},\
  and\ \citenamefont {the ASDEX Upgrade~Team}}]{shao16}%
  \BibitemOpen
  \bibfield  {author} {\bibinfo {author} {\bibfnamefont {L.~M.}\ \bibnamefont
  {Shao}}, \bibinfo {author} {\bibfnamefont {E.}~\bibnamefont {Wolfrum}},
  \bibinfo {author} {\bibfnamefont {F.}~\bibnamefont {Ryter}}, \bibinfo
  {author} {\bibfnamefont {G.}~\bibnamefont {Birkenmeier}}, \bibinfo {author}
  {\bibfnamefont {F.~M.}\ \bibnamefont {Laggner}}, \bibinfo {author}
  {\bibfnamefont {E.}~\bibnamefont {Viezzer}}, \bibinfo {author} {\bibfnamefont
  {R.}~\bibnamefont {Fischer}}, \bibinfo {author} {\bibfnamefont
  {M.}~\bibnamefont {Willensdorfer}}, \bibinfo {author} {\bibfnamefont
  {B.}~\bibnamefont {Kurzan}}, \bibinfo {author} {\bibfnamefont
  {T.}~\bibnamefont {Lunt}}, \ and\ \bibinfo {author} {\bibnamefont {the ASDEX
  Upgrade~Team}},\ }\href {\doibase 10.1088/0741-3335/58/2/025004} {\bibfield
  {journal} {\bibinfo  {journal} {Plasma Physics and Controlled Fusion}\
  }\textbf {\bibinfo {volume} {58}},\ \bibinfo {pages} {025004} (\bibinfo
  {year} {2016})}\BibitemShut {NoStop}%
\bibitem [{\citenamefont {Kobayashi}\ \emph {et~al.}(2016)\citenamefont
  {Kobayashi}, \citenamefont {Itoh}, \citenamefont {Ido}, \citenamefont
  {Kamiya}, \citenamefont {Itoh}, \citenamefont {Miura}, \citenamefont
  {Nagashima}, \citenamefont {Fujisawa}, \citenamefont {Inagaki}, \citenamefont
  {amd},\ and\ \citenamefont {Hoshino}}]{kobayashi16}%
  \BibitemOpen
  \bibfield  {author} {\bibinfo {author} {\bibfnamefont {T.}~\bibnamefont
  {Kobayashi}}, \bibinfo {author} {\bibfnamefont {K.}~\bibnamefont {Itoh}},
  \bibinfo {author} {\bibfnamefont {T.}~\bibnamefont {Ido}}, \bibinfo {author}
  {\bibfnamefont {K.}~\bibnamefont {Kamiya}}, \bibinfo {author} {\bibfnamefont
  {S.-I.}\ \bibnamefont {Itoh}}, \bibinfo {author} {\bibfnamefont
  {Y.}~\bibnamefont {Miura}}, \bibinfo {author} {\bibfnamefont
  {Y.}~\bibnamefont {Nagashima}}, \bibinfo {author} {\bibfnamefont
  {A.}~\bibnamefont {Fujisawa}}, \bibinfo {author} {\bibfnamefont
  {S.}~\bibnamefont {Inagaki}}, \bibinfo {author} {\bibfnamefont {K.~I.}\
  \bibnamefont {amd}}, \ and\ \bibinfo {author} {\bibfnamefont
  {K.}~\bibnamefont {Hoshino}},\ }\href {\doibase 10.1038/srep30720} {\bibfield
   {journal} {\bibinfo  {journal} {Scientific Reports}\ }\textbf {\bibinfo
  {volume} {6}},\ \bibinfo {pages} {30720} (\bibinfo {year}
  {2016})}\BibitemShut {NoStop}%
\bibitem [{\citenamefont {Oberbeck}(1879)}]{oberbeck1879}%
  \BibitemOpen
  \bibfield  {author} {\bibinfo {author} {\bibfnamefont {A.}~\bibnamefont
  {Oberbeck}},\ }\href@noop {} {\bibfield  {journal} {\bibinfo  {journal} {Ann.
  Phys. Chem (Berlin)}\ }\textbf {\bibinfo {volume} {7}},\ \bibinfo {pages}
  {271} (\bibinfo {year} {1879})}\BibitemShut {NoStop}%
\bibitem [{\citenamefont {Boussinesq}(1903)}]{boussinesq03}%
  \BibitemOpen
  \bibfield  {author} {\bibinfo {author} {\bibfnamefont {J.}~\bibnamefont
  {Boussinesq}},\ }\href@noop {} {\bibfield  {journal} {\bibinfo  {journal}
  {\textit{Theorie Analytique De La Chaleur}, Vol. 2}\ } (\bibinfo {year}
  {Gauthier-Villars, Paris, 1903})}\BibitemShut {NoStop}%
\bibitem [{\citenamefont {Xie}\ and\ \citenamefont {Xiao}(2015)}]{xie15}%
  \BibitemOpen
  \bibfield  {author} {\bibinfo {author} {\bibfnamefont {H.~S.}\ \bibnamefont
  {Xie}}\ and\ \bibinfo {author} {\bibfnamefont {Y.}~\bibnamefont {Xiao}},\
  }\href {\doibase 10.1063/1.4931072} {\bibfield  {journal} {\bibinfo
  {journal} {Physics of Plasmas}\ }\textbf {\bibinfo {volume} {22}},\ \bibinfo
  {pages} {090703} (\bibinfo {year} {2015})}\BibitemShut {NoStop}%
\bibitem [{\citenamefont {Xie}, \citenamefont {Xiao},\ and\ \citenamefont
  {Lin}(2017)}]{xie17}%
  \BibitemOpen
  \bibfield  {author} {\bibinfo {author} {\bibfnamefont {H.~S.}\ \bibnamefont
  {Xie}}, \bibinfo {author} {\bibfnamefont {Y.}~\bibnamefont {Xiao}}, \ and\
  \bibinfo {author} {\bibfnamefont {Z.}~\bibnamefont {Lin}},\ }\href {\doibase
  10.1103/PhysRevLett.118.095001} {\bibfield  {journal} {\bibinfo  {journal}
  {Phys. Rev. Lett.}\ }\textbf {\bibinfo {volume} {118}},\ \bibinfo {pages}
  {095001} (\bibinfo {year} {2017})}\BibitemShut {NoStop}%
\bibitem [{\citenamefont {Chen}\ and\ \citenamefont {Chen}(2018)}]{chen18}%
  \BibitemOpen
  \bibfield  {author} {\bibinfo {author} {\bibfnamefont {H.}~\bibnamefont
  {Chen}}\ and\ \bibinfo {author} {\bibfnamefont {L.}~\bibnamefont {Chen}},\
  }\href {\doibase 10.1063/1.5000281} {\bibfield  {journal} {\bibinfo
  {journal} {Physics of Plasmas}\ }\textbf {\bibinfo {volume} {25}},\ \bibinfo
  {pages} {014502} (\bibinfo {year} {2018})}\BibitemShut {NoStop}%
\bibitem [{\citenamefont {Mattor}\ and\ \citenamefont
  {Diamond}(1994)}]{mattor94}%
  \BibitemOpen
  \bibfield  {author} {\bibinfo {author} {\bibfnamefont {N.}~\bibnamefont
  {Mattor}}\ and\ \bibinfo {author} {\bibfnamefont {P.~H.}\ \bibnamefont
  {Diamond}},\ }\href {\doibase 10.1103/PhysRevLett.72.486} {\bibfield
  {journal} {\bibinfo  {journal} {Phys. Rev. Lett.}\ }\textbf {\bibinfo
  {volume} {72}},\ \bibinfo {pages} {486} (\bibinfo {year} {1994})}\BibitemShut
  {NoStop}%
\bibitem [{\citenamefont {Zhang}\ and\ \citenamefont
  {Krasheninnikov}(2017)}]{zhang17}%
  \BibitemOpen
  \bibfield  {author} {\bibinfo {author} {\bibfnamefont {Y.}~\bibnamefont
  {Zhang}}\ and\ \bibinfo {author} {\bibfnamefont {S.~I.}\ \bibnamefont
  {Krasheninnikov}},\ }\href {\doibase 10.1063/1.4994833} {\bibfield  {journal}
  {\bibinfo  {journal} {Physics of Plasmas}\ }\textbf {\bibinfo {volume}
  {24}},\ \bibinfo {pages} {092313} (\bibinfo {year} {2017})}\BibitemShut
  {NoStop}%
\bibitem [{\citenamefont {Madsen}(2013)}]{madsen13}%
  \BibitemOpen
  \bibfield  {author} {\bibinfo {author} {\bibfnamefont {J.}~\bibnamefont
  {Madsen}},\ }\href {\doibase 10.1063/1.4813241} {\bibfield  {journal}
  {\bibinfo  {journal} {Physics of Plasmas}\ }\textbf {\bibinfo {volume}
  {20}},\ \bibinfo {pages} {072301} (\bibinfo {year} {2013})}\BibitemShut
  {NoStop}%
\bibitem [{\citenamefont {Brizard}\ and\ \citenamefont
  {Hahm}(2007)}]{brizard07}%
  \BibitemOpen
  \bibfield  {author} {\bibinfo {author} {\bibfnamefont {A.~J.}\ \bibnamefont
  {Brizard}}\ and\ \bibinfo {author} {\bibfnamefont {T.~S.}\ \bibnamefont
  {Hahm}},\ }\href {\doibase 10.1103/RevModPhys.79.421} {\bibfield  {journal}
  {\bibinfo  {journal} {Review of Modern Physics}\ }\textbf {\bibinfo {volume}
  {79}},\ \bibinfo {pages} {421} (\bibinfo {year} {2007})}\BibitemShut
  {NoStop}%
\bibitem [{\citenamefont {{Spitzer}}(1956)}]{spitzer56}%
  \BibitemOpen
  \bibfield  {author} {\bibinfo {author} {\bibfnamefont {L.}~\bibnamefont
  {{Spitzer}}},\ }\href@noop {} {\emph {\bibinfo {title} {{Physics of Fully
  Ionized Gases}}}}\ (\bibinfo  {publisher} {Interscience Publishers, New
  York},\ \bibinfo {year} {1956})\BibitemShut {NoStop}%
\bibitem [{\citenamefont {Lingam}\ \emph {et~al.}(2017)\citenamefont {Lingam},
  \citenamefont {Hirvijoki}, \citenamefont {Pfefferlé}, \citenamefont
  {Comisso},\ and\ \citenamefont {Bhattacharjee}}]{lingam17}%
  \BibitemOpen
  \bibfield  {author} {\bibinfo {author} {\bibfnamefont {M.}~\bibnamefont
  {Lingam}}, \bibinfo {author} {\bibfnamefont {E.}~\bibnamefont {Hirvijoki}},
  \bibinfo {author} {\bibfnamefont {D.}~\bibnamefont {Pfefferlé}}, \bibinfo
  {author} {\bibfnamefont {L.}~\bibnamefont {Comisso}}, \ and\ \bibinfo
  {author} {\bibfnamefont {A.}~\bibnamefont {Bhattacharjee}},\ }\href {\doibase
  10.1063/1.4980838} {\bibfield  {journal} {\bibinfo  {journal} {Physics of
  Plasmas}\ }\textbf {\bibinfo {volume} {24}},\ \bibinfo {pages} {042120}
  (\bibinfo {year} {2017})}\BibitemShut {NoStop}%
\bibitem [{\citenamefont {Wesson}(2007)}]{wesson07}%
  \BibitemOpen
  \bibfield  {author} {\bibinfo {author} {\bibfnamefont {J.}~\bibnamefont
  {Wesson}},\ }\href@noop {} {\emph {\bibinfo {title} {Tokamaks}}}\ (\bibinfo
  {publisher} {Oxford University Press, Third Edition},\ \bibinfo {year}
  {2007})\BibitemShut {NoStop}%
\bibitem [{\citenamefont {Held}\ \emph {et~al.}(2018)\citenamefont {Held},
  \citenamefont {Wiesenberger}, \citenamefont {Kube},\ and\ \citenamefont
  {Kendl}}]{held18}%
  \BibitemOpen
  \bibfield  {author} {\bibinfo {author} {\bibfnamefont {M.}~\bibnamefont
  {Held}}, \bibinfo {author} {\bibfnamefont {M.}~\bibnamefont {Wiesenberger}},
  \bibinfo {author} {\bibfnamefont {R.}~\bibnamefont {Kube}}, \ and\ \bibinfo
  {author} {\bibfnamefont {A.}~\bibnamefont {Kendl}},\ }\href {\doibase
  10.1088/1741-4326/aad28e} {\bibfield  {journal} {\bibinfo  {journal} {Nuclear
  Fusion}\ } (\bibinfo {year} {2018}),\ 10.1088/1741-4326/aad28e}\BibitemShut
  {NoStop}%
\bibitem [{\citenamefont {Camargo}, \citenamefont {Biskamp},\ and\
  \citenamefont {Scott}(1995)}]{camargo95}%
  \BibitemOpen
  \bibfield  {author} {\bibinfo {author} {\bibfnamefont {S.~J.}\ \bibnamefont
  {Camargo}}, \bibinfo {author} {\bibfnamefont {D.}~\bibnamefont {Biskamp}}, \
  and\ \bibinfo {author} {\bibfnamefont {B.}~\bibnamefont {Scott}},\ }\href
  {\doibase 10.1063/1.871116} {\bibfield  {journal} {\bibinfo  {journal}
  {Physics of Plasmas}\ }\textbf {\bibinfo {volume} {2}},\ \bibinfo {pages}
  {48} (\bibinfo {year} {1995})}\BibitemShut {NoStop}%
\bibitem [{\citenamefont {Numata}, \citenamefont {Ball},\ and\ \citenamefont
  {Dewar}(2007)}]{numata07}%
  \BibitemOpen
  \bibfield  {author} {\bibinfo {author} {\bibfnamefont {R.}~\bibnamefont
  {Numata}}, \bibinfo {author} {\bibfnamefont {R.}~\bibnamefont {Ball}}, \ and\
  \bibinfo {author} {\bibfnamefont {R.~L.}\ \bibnamefont {Dewar}},\ }\href
  {\doibase 10.1063/1.2796106} {\bibfield  {journal} {\bibinfo  {journal}
  {Physics of Plasmas}\ }\textbf {\bibinfo {volume} {14}},\ \bibinfo {pages}
  {102312} (\bibinfo {year} {2007})}\BibitemShut {NoStop}%
\bibitem [{\citenamefont {Trefethen}\ and\ \citenamefont
  {Embree}(2005)}]{trefethen05}%
  \BibitemOpen
  \bibfield  {author} {\bibinfo {author} {\bibfnamefont {L.~N.}\ \bibnamefont
  {Trefethen}}\ and\ \bibinfo {author} {\bibfnamefont {M.}~\bibnamefont
  {Embree}},\ }\href@noop {} {\emph {\bibinfo {title} {{Spectra and
  Pseudospectra: The Behavior of Nonnormal Matrices and Operators}}}}\
  (\bibinfo  {publisher} {{Princeton University Press}},\ \bibinfo {year}
  {2005})\BibitemShut {NoStop}%
\bibitem [{\citenamefont {Wiesenberger}\ and\ \citenamefont
  {Held}(2018)}]{feltor41}%
  \BibitemOpen
  \bibfield  {author} {\bibinfo {author} {\bibfnamefont {M.}~\bibnamefont
  {Wiesenberger}}\ and\ \bibinfo {author} {\bibfnamefont {M.}~\bibnamefont
  {Held}},\ }\href {\doibase 10.5281/zenodo.1207806} {\enquote {\bibinfo
  {title} {{FELTOR v4.1}},}\ } (\bibinfo {year} {2018})\BibitemShut {NoStop}%
\bibitem [{\citenamefont {Schmid}(2007)}]{schmid07}%
  \BibitemOpen
  \bibfield  {author} {\bibinfo {author} {\bibfnamefont {P.~J.}\ \bibnamefont
  {Schmid}},\ }\href {\doibase 10.1146/annurev.fluid.38.050304.092139}
  {\bibfield  {journal} {\bibinfo  {journal} {Annual Review of Fluid
  Mechanics}\ }\textbf {\bibinfo {volume} {39}},\ \bibinfo {pages} {129}
  (\bibinfo {year} {2007})}\BibitemShut {NoStop}%
\bibitem [{\citenamefont {Held}\ and\ \citenamefont {Kendl}(2015)}]{held15}%
  \BibitemOpen
  \bibfield  {author} {\bibinfo {author} {\bibfnamefont {M.}~\bibnamefont
  {Held}}\ and\ \bibinfo {author} {\bibfnamefont {A.}~\bibnamefont {Kendl}},\
  }\href {\doibase 10.1016/j.jcp.2015.06.018} {\bibfield  {journal} {\bibinfo
  {journal} {Journal of Computational Physics}\ }\textbf {\bibinfo {volume}
  {298}},\ \bibinfo {pages} {622 } (\bibinfo {year} {2015})}\BibitemShut
  {NoStop}%
\bibitem [{\citenamefont {Kendl}(2018)}]{kendl18}%
  \BibitemOpen
  \bibfield  {author} {\bibinfo {author} {\bibfnamefont {A.}~\bibnamefont
  {Kendl}},\ }\href {\doibase 10.1088/1361-6587/aa9f94} {\bibfield  {journal}
  {\bibinfo  {journal} {Plasma Physics and Controlled Fusion}\ }\textbf
  {\bibinfo {volume} {60}},\ \bibinfo {pages} {025017} (\bibinfo {year}
  {2018})}\BibitemShut {NoStop}%
\bibitem [{\citenamefont {Staebler}, \citenamefont {Kinsey},\ and\
  \citenamefont {Waltz}(2005)}]{staebler05}%
  \BibitemOpen
  \bibfield  {author} {\bibinfo {author} {\bibfnamefont {G.~M.}\ \bibnamefont
  {Staebler}}, \bibinfo {author} {\bibfnamefont {J.~E.}\ \bibnamefont
  {Kinsey}}, \ and\ \bibinfo {author} {\bibfnamefont {R.~E.}\ \bibnamefont
  {Waltz}},\ }\href {\doibase 10.1063/1.2044587} {\bibfield  {journal}
  {\bibinfo  {journal} {Physics of Plasmas}\ }\textbf {\bibinfo {volume}
  {12}},\ \bibinfo {pages} {102508} (\bibinfo {year} {2005})}\BibitemShut
  {NoStop}%
\bibitem [{\citenamefont {Bourdelle}\ \emph {et~al.}(2007)\citenamefont
  {Bourdelle}, \citenamefont {Garbet}, \citenamefont {Imbeaux}, \citenamefont
  {Casati}, \citenamefont {Dubuit}, \citenamefont {Guirlet},\ and\
  \citenamefont {Parisot}}]{bourdelle07}%
  \BibitemOpen
  \bibfield  {author} {\bibinfo {author} {\bibfnamefont {C.}~\bibnamefont
  {Bourdelle}}, \bibinfo {author} {\bibfnamefont {X.}~\bibnamefont {Garbet}},
  \bibinfo {author} {\bibfnamefont {F.}~\bibnamefont {Imbeaux}}, \bibinfo
  {author} {\bibfnamefont {A.}~\bibnamefont {Casati}}, \bibinfo {author}
  {\bibfnamefont {N.}~\bibnamefont {Dubuit}}, \bibinfo {author} {\bibfnamefont
  {R.}~\bibnamefont {Guirlet}}, \ and\ \bibinfo {author} {\bibfnamefont
  {T.}~\bibnamefont {Parisot}},\ }\href {\doibase 10.1063/1.2800869} {\bibfield
   {journal} {\bibinfo  {journal} {Physics of Plasmas}\ }\textbf {\bibinfo
  {volume} {14}},\ \bibinfo {pages} {112501} (\bibinfo {year}
  {2007})}\BibitemShut {NoStop}%
\end{thebibliography}%
\bibliographystyle{aipnum4-1.bst}
\end{document}